\documentclass[11pt, reqno]{article}
\usepackage{jheppub}
\usepackage{epsfig}
\usepackage{amssymb}
\usepackage{amsmath}
\usepackage{mathrsfs}
\usepackage{hyperref}
\usepackage{multirow}
\usepackage{subcaption}
\usepackage{comment}

\usepackage{tikz}
\usetikzlibrary{shapes,arrows,automata,positioning,decorations.pathmorphing}

\DeclareMathOperator{\Tr}{Tr}

\usepackage{feynmp}
\newlength{\eqoff}

\usepackage{ifpdf}

\ifpdf
\else
\fi

\usepackage{xkeyval}
\input{picturemacro}

\makeatletter
\DeclareRobustCommand*{\bfseries}{%
  \not@math@alphabet\bfseries\mathbf
  \fontseries\bfdefault\selectfont
  \boldmath
}

\makeatother

\definecolor{labelcolor}{rgb}{0,0,0}

\usepackage[textsize=tiny,textwidth=2.25cm]{todonotes}

\newcommand{\e}{\operatorname{e}}
\newcommand{\de}{\operatorname{d}\!}
\newcommand{\peps}{\varepsilon}

\def\be{\begin{equation}}
\def\ee{\end{equation}}
\def\ba{\begin{eqnarray}}
\def\ea{\end{eqnarray}}
\def\nl{\nonumber\\}
\def\l{\langle}
\def\r{\rangle}

\def\tlambda{\tilde{\lambda}}
\def\dalpha{{\dot{\alpha}}}
\def\dbeta{{\dot{\beta}}}
\def\CA{C_A}
\def\CF{C_F}

\def\MM{\mathcal{M}}
\def\IR{{\textrm{IR}}}
\def\UV{{\textrm{UV}}}
\def\muUV{\mu_{\UV}}
\def\tildeeta{\tilde{\eta}}

\DeclareMathOperator{\im}{Im}
\renewcommand{\Im}{\im}

\def\OO{\mathcal{O}}

\def\FO#1{F_{\OO}\!\big[#1\big]}

\def\gcoll{\gamma^{\rm coll.}}

\newcommand{\fermm}{{\bar{\psi}}}
\newcommand{\fermp}{{\psi}}

\title{Renormalization Group Coefficients and the S-Matrix}

\author[a,b]{Simon Caron-Huot}
\author[b]{and Matthias Wilhelm}
\affiliation[a]{Niels Bohr International Academy and Discovery Center}%, Blegdamsvej 17, Copenhagen 2100 \O{}, Denmark}
\emailAdd{schuot@nbi.dk} 
\affiliation[b]{Niels Bohr Institute, Blegdamsvej 17, Copenhagen 2100 \O{}, Denmark}
\emailAdd{matthias.wilhelm@nbi.ku.dk}
\abstract{%
We show how to use on-shell unitarity methods to calculate renormalization group coefficients such as beta functions and anomalous dimensions.
The central objects are the form factors of composite operators.
Their discontinuities can be calculated via phase-space integrals and are related to corresponding anomalous dimensions.
In particular, we find that the dilatation operator, which measures the anomalous dimensions, is given by minus the phase of the $S$-matrix divided by $\pi$. 
We illustrate our method using several examples from Yang-Mills theory, perturbative QCD and Yukawa theory at one-loop level and beyond.
}
\begin{document}\begin{fmffile}{diagrams}
\fmfcmd{%
thin := 1pt; % dimension of the lines
thick := 2thin;
arrow_len := 3mm;
arrow_ang := 15;
curly_len := 3mm;
dash_len :=0.3; % 'photon' lines
dot_len := 0.75mm; % 'photon' lines
wiggly_len := 2mm; % 'photon' lines
wiggly_slope := 60;
zigzag_len := 2mm;
zigzag_width := 2thick;
decor_size := 5mm;
dot_size := 2thick;
}
\fmfcmd{%
marksize=7mm;
def draw_cut(expr p,a) =
  begingroup
    save t,tip,dma,dmb; pair tip,dma,dmb;
    t=arctime a of p;
    tip =marksize*unitvector direction t of p;
    dma =marksize*unitvector direction t of p rotated -90;
    dmb =marksize*unitvector direction t of p rotated 90;
    linejoin:=beveled;
    drawoptions(dashed dashpattern(on 3bp off 3bp on 3bp));
    draw ((-.5dma.. -.5dmb) shifted point t of p);
    drawoptions();
  endgroup
enddef;
style_def phantom_cut expr p =
    save amid;
    amid=.5*arclength p;
    draw_cut(p, amid);
    draw p;
enddef;
}
\fmfcmd{%
smallmarksize=4mm;
def draw_smallcut(expr p,a) =
  begingroup
    save t,tip,dma,dmb; pair tip,dma,dmb;
    t=arctime a of p;
    tip =smallmarksize*unitvector direction t of p;
    dma =smallmarksize*unitvector direction t of p rotated -90;
    dmb =smallmarksize*unitvector direction t of p rotated 90;
    linejoin:=beveled;
    drawoptions(dashed dashpattern(on 2bp off 2bp on 2bp) withcolor red);
    draw ((-.5dma.. -.5dmb) shifted point t of p);
    drawoptions();
  endgroup
enddef;
style_def phantom_smallcut expr p =
    save amid;
    amid=.5*arclength p;
    draw_smallcut(p, amid);
    draw p;
enddef;
style_def plain_arrow_smallcut expr p =
    cdraw p;
    cfill (arrow p);
    save amid;
    amid=.5*arclength p;
    draw_smallcut(p, amid);
    draw p;
enddef;
}

\fmfcmd{%
style_def plain_ar expr p =
  cdraw p;
  shrink (0.6);
  cfill (arrow p);
  endshrink;
enddef;
style_def plain_rar expr p =
  cdraw p; 
  shrink (0.6);
  cfill (arrow reverse(p));
  endshrink;
enddef;
style_def dashes_ar expr p =
  draw_dashes p;
  shrink (0.6);
  cfill (arrow p);
  endshrink;
enddef;
style_def dashes_rar expr p =
  draw_dashes p;
  shrink (0.6);
  cfill (arrow reverse(p));
  endshrink;
enddef;
style_def dots_ar expr p =
  draw_dots p;
  shrink (0.6);
  cfill (arrow p);
  endshrink;
enddef;
style_def dots_rar expr p =
  draw_dots p;
  shrink (0.6);
  cfill (arrow reverse(p));
  endshrink;
enddef;
}

\fmfcmd{%
style_def phantom_cross expr p =
    save amid,ang;
    amid=.5*length p;
    ang= angle direction amid of p;
    draw ((polycross 4) scaled 8 rotated ang) shifted point amid of p;
enddef;
}

\fmfcmd{%
style_def plain_sarrow expr p =
  cdraw p;
  shrink (0.55); %  shrink (0.6);
  cfill (arrow p);
  endshrink;
enddef;
style_def dashes_sarrow expr p =
  draw_dashes p;
  shrink (0.55);
  cfill (arrow p);
  endshrink;
enddef;
style_def dots_sarrow expr p =
  draw_dots p;
  shrink (0.55);
  cfill (arrow p);
  endshrink;
enddef;
style_def plain_srarrow expr p =
  cdraw p;
  shrink (0.55);
  cfill (arrow (reverse p));
  endshrink;
enddef;
style_def dashes_srarrow expr p =
  draw_dashes p;
  shrink (0.55);
  cfill (arrow (reverse p));
  endshrink;
enddef;
style_def dots_srarrow expr p =
  draw_dots p;
  shrink (0.55);
  cfill (arrow (reverse p));
  endshrink;
enddef;
}

\notoc
\maketitle
\newpage

\section{Introduction}

On-shell approaches play a central role in many state-of-the-art calculations in perturbative gauge theories.
Since only physical degrees of freedom appear on-shell, they enable to build observables in terms of the simplest but meaningful physical building blocks.
This is especially advantageous for massless particles with spin, such as gluons, where the focus on the two physical helicities
removes the need to introduce gauge redundancies, removing at the same time intricate cancellations among large numbers of Feynman diagrams.

In on-shell approaches, the Lagrangian and Feynman rules of a theory tend to occupy a secondary role, if any.
It is therefore crucial to develop a conceptual understanding, directly in the language that is used in calculations,
of the phenomena that are traditionally understood from the Lagrangian.  
In this paper, we discuss a direct connection
between the high-energy behavior of the $S$-matrix of a theory and the running of coupling constants and renormalization of local operators.
We will build on recent developments in the context of
the dilatation operator in  $\mathcal{N}=4$ super Yang-Mills (SYM) \cite{Zwiebel:2011bx,Wilhelm:2014qua,Nandan:2014oga,Koster:2014fva,Brandhuber:2014pta,Brandhuber:2015boa,Loebbert:2015ova,Brandhuber:2016fni} and other work based on generalized unitarity \cite{ArkaniHamed:2008gz,Huang:2012aq,Cheung:2015aba}, which we will extend to arbitrary weakly coupled field theories.

Our main physical idea will be the notion that large logarithms signaling the running of couplings originate from states which propagate over a ``long distance'' in an appropriate metric,
making them effectively on-shell.
Quantitatively, we will consider form factors, which are matrix elements between an operator and on-shell states:
\begin{equation}
\FO{p_1,\ldots,p_n;\mu} \equiv \l p_1,\ldots, p_n | \OO|0\r\,, \label{form_factor}
\end{equation}
where $\mu$ is the renormalization scale.
Such form factors figure prominently in effective-theory descriptions of weak processes including Higgs production and decay, see e.g.~\cite{Schmidt:1997wr,Anastasiou:2015ema}.
They convert the scale dependence of the local operator $\OO$ into a physically measurable energy dependence of its decay products.
The key fact for us will be that the energy dependence and phase are tied to each other, as can be seen from the imaginary part
acquired by the logarithms for timelike momentum invariants ($p^2>0$)
due to Feynman's prescription $p^2\to p^2+i0$:
\begin{equation}
\log \left(\frac{-p^2}{\mu^2}\right) \equiv
\log \left(\frac{|p^2|}{\mu^2}\right)-i\pi \quad\Rightarrow\quad
p^\mu \frac{\partial}{\partial p^\mu} \log \left(\frac{-p^2}{\mu^2}\right) =  -\frac{2}{\pi} \Im \log \left(\frac{-p^2}{\mu^2}\right)\,.
\label{logs_imaginary_part}
\end{equation}
This is interesting because, as understood from conventional unitarity and the optical theorem,
imaginary parts originate physically from the long time propagation of intermediate on-shell states.
This suggests that the scale dependence of a process can be understood directly from the
propagation of on-shell particles.
In this paper, we propose a precise quantitative relationship, which we will verify in a number of classic examples.

This paper is organized as follows.
In section \ref{sec: S matrix as dilatation operator}, we expand on the ideas sketched above, deriving a relation between the $S$-matrix and the dilatation operator, and we set up our notations.
In section \ref{sec: one-loop application}, we apply these ideas at one-loop level.
We calculate the $\beta$-functions and anomalous dimensions of various composite operators in pure Yang-Mills, perturbative QCD and $\mathcal{N}=4$ SYM.
In section \ref{sec: towards higher loops}, we extend our study to several features that appear at higher loop orders, in particular the mixing of operators of different lengths.
We conclude with a summary of our results and an outlook on future directions in section \ref{sec: Summary}.

\section{The \texorpdfstring{$S$}{S}-matrix and the dilatation operator}
\label{sec: S matrix as dilatation operator}

In this section, we derive a concrete formula, eq.~(\ref{cute_eigenvalue_equation}),
which instantiates the above general ideas, and we set up the notations we will use to test it.

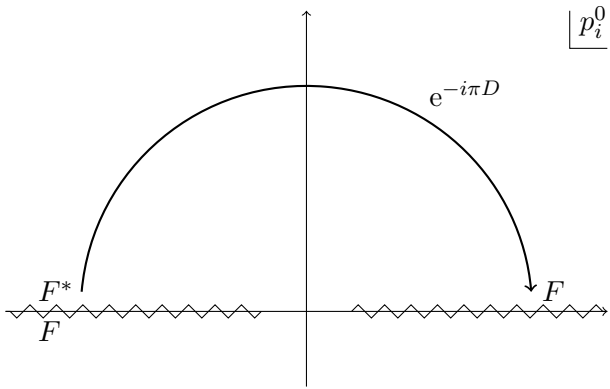
\begin{figure}[htbp]
\begin{center} 
\begin{tikzpicture}
	[	axis/.style={->,black},
		continuation/.style={->,black,thick},
		line/.style={black,thick},
		discontinuity/.style={black}]
	
	%draw the axes
	\draw[axis] (-4,0) -- (4,0) node[anchor=west]{};
	\draw[axis] (0,-1) -- (0,4) node[anchor=south]{};
	%draw circle
	\draw[continuation] ++(175:3) arc (175:5:3);
	%draw discontinuity
	\draw[discontinuity,decoration={zigzag},decorate] (-0.6,0) -- (-4,0);
	\draw[discontinuity,decoration={zigzag},decorate] (0.6,0) -- (4,0);
	%Labels
	\node[anchor=south west] at (3,0) {$F$};
	\node[anchor=north west] at (-3.7,0) {$F$};
	\node[anchor=south west] at (-3.7,0) {$F^*$};
	\node[anchor=south west] at (60:3) {$\e^{-i\pi D}$};
	\draw (4,3.5) -- (3.5,3.5) -- (3.5,4.0);
	\node[anchor=south west] at (3.5,3.5) {$p_i^0$};
\end{tikzpicture}
\end{center} 
\caption{Analytic continuation from the complex conjugate using a complex scale transformation.}
\label{fig: half-circle}
\end{figure}

The main first step is to connect the phase and energy dependence of form factors. This connection stems from analyticity.
The trick is to use a complex scale transformation to relate a form factor to its complex conjugate.
We start from a kinematic configuration where all momenta $p_i$ are outgoing,
so that all Mandelstam invariants are positive (timelike): $s_{ij\ldots k}=(p_i+p_j+\ldots+p_k)^2>0$.
The form factor is not real because the Feynman prescription adds a small positive imaginary part to all invariants: $s_{I}\mapsto s_{I}+i0$.
But it can be related to its conjugate by an analytic continuation in which all the invariants are rotated along a large circle in the complex plane,
with a common phase, as illustrated in fig.~\ref{fig: half-circle}.
Such a rotation is generated by the dilatation operator $D$:
\begin{equation}
 F(p_1,\ldots,p_n) \to F(p_1\e^{i\alpha},\ldots,p_n\e^{i\alpha}) = \e^{i\alpha D} F(p_1,\ldots,p_n)\,,\quad\mbox{where}\quad D\equiv \sum_i p_i^\mu \frac{\partial}{\partial p_i^\mu}\,.
\end{equation}
We do not expect any singularity until the angle reaches $\pi$, where all energies are reversed.
(This is easily proved in perturbation theory, where the Feynman parameter representation contains denominators
of the form $\big(\sum_j c_j m_j^2-\sum_J c_J s_{J}-i0\big)$ with all $c_j,c_J$ positive. Taking all $s_J$ to have the same phase $\e^{2i\alpha}$, the first singularity is at $\alpha=\pi$.)
At this point, the invariants are back to the original ones but on the ``wrong'' side of the cut,
giving the conjugate form factor. Thus,
\begin{equation}
 F = \e^{-i\pi D}F^*\,, \label{continuation_from_D}
\end{equation}
where $F^*$ is the form factor computed using anti-time-ordered propagators.

The second fundamental equation we will need is a version of the optical theorem. The conventional optical theorem
expresses unitarity of the $S$-matrix: $SS^\dagger=1$,
where the product contains a phase-space integral over intermediate $n$-particle states summed over all $n$.
Formally using the physical interpretation of a form factor as a small perturbation to the $S$-matrix,
$\delta S=i\mathcal{F}$, using the calligraphic font here to distinguish the operator $\mathcal{F}$ from its matrix elements to outgoing states $F$,
unitarity becomes $\mathcal{F}= S \mathcal{F}^\dagger S$. For vacuum initial states, this reduces to
\begin{equation}
 F = S F^*\,.
\end{equation}
In this note, we will mostly rely on the imaginary part of this relation to one-loop order, which is easily verified from the Cutkowski rules.
The diagrams which contribute to the product $SF^*$ originate by drawing a cut through form factor diagrams, as depicted for example in fig.~\ref{fig: one-loop double cut} below. The massless scattering amplitudes contained in $S$ then join the cut to the final states.
We note that the other side of the cut involves a complex conjugate amplitude, as is typically the case for Cutkowski rules.

Combining the two relations above gives
\begin{equation}
\e^{-i\pi D}F^*=S F^*\,. \label{cute_eigenvalue_equation}
\end{equation}
This will be the central equation in this paper. We will read it as follows:
{\it the dilatation operator is minus the phase of the $S$-matrix, divided by $\pi$.}\footnote{Strictly speaking, we are omitting a CPT transformation here,
whose necessity can be seen for example using the commutation relation with the Hamiltonian $H$. We thank Amit Sever for this observation.}

The dilatation operator is of course closely related to renormalization group evolution.  Indeed, at high energies,
by dimensional analysis, $F$ can depend only on dimensionless ratios $s_{I}/\mu^2$, and therefore
$D\simeq-\mu\partial_\mu$. Starting from the renormalization group equation
\begin{equation}
 \left[\mu\partial_\mu +\beta(g^2)\frac{\partial}{\partial g^2} + \gamma_\OO - \gamma_{\rm IR} \right]F = 0\,, \label{eq: RG equation mu}
\end{equation}
one hence obtains
\begin{equation}
DF=\left(\gamma_\OO - \gamma_{\rm IR}+\beta(g^2)\frac{\partial}{\partial g^2}\right)F\,.  \label{rg_equation}
\end{equation}
It will be important that ultraviolet and infrared divergences both contribute to the energy dependence of form factors. Their relative sign is simply a convention
which ensures that the naturally large logarithms $\log(\mu_{\rm UV}^2/p^2)$ and $\log(p^2/\mu_{\rm IR}^2)$ come with the same sign
when their renormalization scales are treated independently.
Logarithms of momentum-independent masses will be discussed briefly in section \ref{eq: comment on masses} but do not fundamentally affect the discussion.

Inserting (the complex conjugate of) (\ref{rg_equation}) into (\ref{cute_eigenvalue_equation})
yields a relation between the renormalization group coefficients $\gamma_\OO$, $\gamma_{\rm IR}$, $\beta$ and the S-matrix.
Let us focus on the leading approximation to this otherwise exact relation.
It is useful to restrict to so-called minimal form factors, which are non-vanishing in the free-theory limit.
The $\beta$-function term can then be neglected. Writing $S=1+i\mathcal{M}$ and
inserting (\ref{rg_equation}) into (\ref{cute_eigenvalue_equation}) then gives to leading non-trivial order:
\begin{equation}
 \left(\gamma_\OO^{(1)} - \gamma_{\rm IR}^{(1)}\right) \l p_1,\ldots,p_n| \OO|0\r^{(0)}
 = -\frac{1}{\pi} \l p_1,\ldots,p_n| \mathcal{M}\otimes \OO|0\r^{(0)}\,, \label{eq: gamma UV plus IR}
\end{equation}
where $\mathcal{M}$ is the tree-level $2{\to}2$ $S$-matrix, and
the convolution, to be defined shortly, represents the phase-space integral over intermediate two-particle states
in the product $\mathcal{M} F^*$. 
Note that we have dropped the complex conjugation sign, as the tree-level form factors are naturally real. 

In order to use the above equation to extract anomalous dimensions, the infrared contributions must be subtracted.
The key fact is that these depend only on the external particles but not on $\OO$. This makes it possible to construct infrared-safe ratios.
This is particularly simple in the special case $n=2$, where one can put the stress-tensor in the denominator.
It has vanishing anomalous dimension in any theory. Ignoring again the $\beta$-function, this gives
\begin{equation}
 \gamma_{\OO} = D \log \frac{\langle p_1,p_2|\OO|0\rangle}{\langle p_1,p_2|T^{\mu\nu}|0\rangle}
 =-\frac{1}{\pi} 2\Im \log \frac{\langle p_1,p_2|\OO|0\rangle}{\langle p_1,p_2|T^{\mu\nu}|0\rangle}\,, \label{eq: IR safe ratio}
\end{equation}
which gives rise to the more practical one-loop equation
\begin{equation}
 \gamma_\OO^{(1)} \l p_1,p_2| \OO|0\r^{(0)}
 = -\frac{1}{\pi} \l p_1,p_2| \mathcal{M}\otimes \OO|0\r^{(0)}+\frac{1}{\pi} \l p_1,p_2| \OO|0\r^{(0)} \frac{\l p_1,p_2|\mathcal{M}\otimes T^{\mu\nu}|0\r^{(0)}}{\l p_1,p_2|T^{\mu\nu}|0\r^{(0)}}\,.
\label{eq: gammaO practical}
\end{equation}
This equation is new. 
Multiple examples and applications will be given in sections \ref{sec: one-loop application} and \ref{sec: towards higher loops}.
Note that the ratio in \eqref{eq: gammaO practical} does not depend on the indices on $T^{\mu\nu}$ because the infrared divergences are blind to these.
At higher loops and in the presence of a $\beta$-function, the imaginary part of the above logarithm is still useful and detects
the anomalous dimensions and coupling dependence of the form factor averaged over the half-circle of fig.~\ref{fig: half-circle}.

The anomalous dimensions of marginal and relevant operators are of particular physical interest due to
their relation to the $\beta$-functions of corresponding running couplings.
For example, in Yang-Mills theory,
the anomalous dimension of the Lagrangian density is a derivative of the $\beta$-function \cite{KlubergStern:1974rs,Grinstein:1988wz}:
\begin{equation}
 \gamma_{\mathcal{L}} = g^2 \frac{\partial}{\partial g^2} \left(\frac{\beta(g^2)}{g^2}\right)\,. \label{eq:gamma vs beta function}
\end{equation}
The two are therefore essentially equivalent, making it possible
to use the preceding formulas to obtain $\beta$-functions. The multi-coupling case will be discussed further in section \ref{sec: towards higher loops}.

Note that the arguments above are valid in any space-time dimension.
In the following, we will restrict ourselves to four dimensions though.

\subsection{Notations: form factors and spinor-helicity variables}

Form factors provide a map between on-shell states and local operators.
In a free theory, they are just polynomials in the momenta. For example, for a free scalar
\begin{equation}
 i^n \langle 1_\phi | \partial^{\mu_1}\cdots \partial^{\mu_n} \phi|0\rangle = p_1^{\mu_1}\cdots p_1^{\mu_n}\,.
\end{equation}
In general, for final state with multiple particles, there is a one-to-one correspondence between such polynomials and local operators modulo equations of motion.
Note that we use an abbreviated notation where the bra $\l 1_h|$ denotes a particle of type $h$ with momentum $p_1$.

When dealing with particles with spin, it is useful to use variables which can absorb the phase ambiguities
of their polarization vectors and spinors. In four dimensions, this is nicely achieved by the so-called spinor-helicity variables.
These are defined by splitting a null four-momentum into two Weyl spinors:
\begin{equation}
 p_j^{\alpha\dalpha} \equiv p_j^\mu \sigma_\mu^{\alpha\dalpha}= \lambda_j^\alpha \tilde\lambda_j^{\dalpha} \,,
 \label{eq: spinor helicity}
\end{equation}
where $(\sigma_\mu)^{\alpha\dalpha}$ are the four-dimensional ($2\times 2$) Pauli matrices.
The two helicity polarizations of a gluon can be parametrized explicitly in terms of the spinors, see for example \cite{Henn:2014yza}.
The important fact is that the physics is invariant if spinors and antispinors are rotated by opposite phases,
provided the external states are simultaneously rotated according to their helicity:
\begin{equation}
 \lambda_j\to \lambda_j \e^{i\alpha_j}\,,\quad
 \tilde\lambda_j\to \tilde\lambda_j \e^{-i\alpha_j}\,,\quad
\langle j_-| \to \e^{2i\alpha_j}\langle j_-|\,,\quad
 \langle j_+| \to \e^{-2i\alpha_j}\langle j_+|\,. \label{eq:little group}
\end{equation}
This is called little-group scaling because the same phases would arise from a rotation along the propagation axis of particle $j$.
Thus, form factors are polynomials in the spinor-helicity variables with a specific little-group weight for each particle.
This fixes the form of form factors for the self-dual and anti-self-dual parts of the field strength and fermion fields, 
\begin{equation}\begin{aligned}
 \langle 1_-|F^{\alpha\beta}|0\rangle \equiv \lambda_1^{\alpha}\lambda_1^{\beta}\,,\quad
 \langle 1_+|\bar{F}^{\dot\alpha\dot\beta}|0\rangle \equiv \tilde{\lambda}_1^{\dalpha}\tilde{\lambda}_1^{\dbeta}\,,\quad
 \langle 1_{\fermm}| \psi^{\alpha}|0\rangle \equiv \lambda_1^{\alpha}\,,\quad
 \langle 1_{\fermp}| \bar{\psi}^{\dot\alpha}|0\rangle \equiv \tilde{\lambda}_1^{\dalpha}\,,
\end{aligned}\end{equation}
where the state $\langle 1_{\fermp}|$ is a Weyl fermion of positive helicity.
We follow conventions where the basic Lorentz invariant combinations are the brackets
\begin{equation}
 s_{ij}=2p_i{\cdot}p_j = \langle i\,j\rangle[j\,i]\,,\quad\mbox{where}\quad
 \langle i\,j\rangle \equiv \epsilon_{\alpha\beta}\lambda_i^\alpha\lambda_j^\beta\,,\quad
 [i\,j] \equiv \epsilon_{\dot\alpha\dot\beta}\tilde\lambda_i^{\dot\alpha}\tilde\lambda_j^{\dot\beta}\,,
\end{equation}
with the Mandelstam invariant $s_{ij}>0$ when the invariant is timelike, as is the case for two outgoing particles.
For outgoing momenta, there is the complex conjugation relation $\lambda_i=(\tilde\lambda_i)^*$.

Like its name suggests, the $S$-matrix $S=1+i\mathcal{M}$ is an operator,
which in particular can act on the polynomial states produced by minimal form factors. This action, which we denote as a convolution,
is simply the on-shell phase-space integral:
\ba
 \l 12| \mathcal{M}\otimes F|0\r^{(0)} \equiv \frac{1}{16\pi} \sum_{h_{1'},h_{2'}} 
 \int \frac{\de\Omega}{4\pi} \l 12| \mathcal{M}| 1'_{h_{1'}}2'_{h_{2'}}\r^{(0)}
 \l 1'_{h_{1'}}2'_{h_{2'}}| F|0\r^{(0)}\,, \label{eq: phase space}
\ea
where the sum is over all intermediate helicity states.
The following elegant phase-space parametrization using spinors will be useful: one simply rotates the spinors as \cite{Zwiebel:2011bx}
\begin{equation}
 \left(\begin{array}{c} \lambda_1'\\ \lambda_2'\end{array}\right)=
 \left(\begin{array}{cc} \cos\theta & -\sin\theta \e^{i\phi}\\\sin\theta \e^{-i\phi}&\cos\theta\end{array}\right)
 \left(\begin{array}{c} \lambda_1\\ \lambda_2\end{array}\right)\,, \label{rotation}
\end{equation}
together with the complex conjugate rotation for the conjugate spinors $\tilde{\lambda}_1'$ and $\tilde{\lambda}_2'$.
It is easy to verify that $p_1'+p_2'=p_1+p_2$. 
In a center-of-mass frame where $p_1$ and $p_2$ are back-to-back along the $z$-axis, this reduces to a standard parametrization of spinors in terms of
polar half-angle $\theta$ and azimuthal angle $\phi$.  The advantage is that, being covariant, this can be used in any frame.
The integration measure is simply
\begin{equation}
\label{eq: one-loop measure}
\int \frac{\de\Omega}{4\pi} \equiv
\int \frac{\de\phi}{2\pi}\int_0^{\frac{\pi}{2}}2\cos\theta\sin\theta \de\theta\,.
\end{equation}

Finally, following general practice in the amplitudes community, we will use crossing symmetry liberally
and often express $S$-matrix elements in a notation where momenta and other quantum numbers are outgoing:
\begin{equation}
(-1)^{n_{\fermm}}
 \l 1_{h_1}2_{h_2}| \mathcal{M} | 3_{h_3}4_{h_4}\r  \equiv \l 1_{h_1}2_{h_2}\bar{4}_{-h_4}\bar{3}_{-h_3}|\mathcal{M}|0\r\equiv \mathcal{M}_{1_{h_1}2_{h_2}\bar{4}_{-h_4}\bar{3}_{-h_3}}\,,
 \label{eq: crossing}
\end{equation}
where $\bar{j}$ means minus the momentum $p_j$ (with $\lambda_{\bar{j}}=\lambda_j, \tilde{\lambda}_{\bar{j}}=-\tilde{\lambda}_{j}$).
Reversing the order of fields upon crossing is a useful convention which ensures the proper minus signs for fermion loops.
An additional minus sign counts the number of negative-helicity fermions $\fermm$ in the initial state.

\section{Application: Yang-Mills theory at one-loop}
\label{sec: one-loop application}

To compute all one-loop anomalous dimensions in Yang-Mills theory, the major ingredient will be
the on-shell four-gluon amplitude, given by the famous Parke-Taylor expression:
\begin{equation}
\label{eq: Parke-Taylor formula}
 \mathcal{M}^{abcd}_{1^-2^-3^+4^+} = -2g^2 \l 12\r^4 \left[ \frac{f^{abe}f^{cde}}{\l12\r\l23\r\l34\r\l41\r}+\frac{f^{ace}f^{bde}}{\l13\r\l32\r\l24\r\l41\r}\right]\,.
\end{equation}
For other helicity choices, one simply replaces $\l12\r^4$ by $\l ij\r^4$, where $i$ and $j$ are the two negative-helicity gluons;
the four-gluon tree amplitude vanishes if there are not exactly two negative-helicity gluons.
We will mostly need the case where the initial state is a color-singlet gluon pair, in which case the formula simplifies as the first term vanishes:
\begin{equation}
 \mathcal{M}^{abcd}_{1^-2^-3^+4^+} \delta^{cd}= -2g^2\CA \delta^{ab} \frac{\l 12\r^4}{\l 13\r\l32\r\l24\r\l41\r}\,. \label{eq: Parke-Taylor formula_trace}
\end{equation}
Here, $\CA$ denotes the Casimir in the adjoint representation, which is $N_c$ for gauge group SU(N${}_c$).
Before using it, let us briefly comment on various ways to obtain eq.~(\ref{eq: Parke-Taylor formula_trace}), which of course include
direct Feynman diagram calculation \cite{Srednicki:2007qs,Elvang:2013cua}. It is also a special case of the celebrated MHV $n$-point amplitude,
now understood from a large number of viewpoints including Berends-Giele \cite{Berends:1987me} and BCFW recursion \cite{Britto:2004ap,Britto:2005fq}, properties of self-dual Yang-Mills \cite{Rosly:1996vr}, the twistor string \cite{Witten:2003nn}, etc.
In fact, the above formula is a direct consequence of basic physical principles, specifically its little-group properties and classical small-angle limits.
The key point is that the little-group scaling (\ref{eq:little group}) implies that the amplitude can be written as $\frac{\l 34\r^2}{\l 12\r^2}$ times a rational function $G(s,t,u)$. Since a tree amplitude cannot have a squared denominator such as $1/\l 12\r^2$, $G$ needs to be proportional to $s=\l12\r[21]$, and since it needs to be dimensionless and only massless poles can appear in its denominator,
the most general possibility is $G=c_1 \frac{s}{t} + c_2 \frac{s}{u}$. In the small-angle limit $t\to 0$,
the amplitude has to reproduce the Coulomb-like attractive potential $\MM\to -2g^2f^{a_1a_4b}f^{a_2a_3b}\tfrac{s}{t}$,
and similarly at $u\to 0$, which fixes $c_1=c_2=-2g^2\CA$. This reproduces eq.~(\ref{eq: Parke-Taylor formula_trace}) using spinor identities.
The absence of polynomial ambiguities for massless particles with spin is a generic consequence of little-group scaling \cite{Weinberg:1965rz,Benincasa:2007xk}.

Plugging in the explicit values for the rotated spinors in eq.~(\ref{rotation}),
\begin{equation}
\l 1'2'\r=\l12\r,\quad \l12'\r=\l1'2\r=\l12\r\cos\theta,\quad \l1'1\r=\l12\r\sin\theta \e^{i\phi},\quad \l2'2\r=\l12\r\sin\theta \e^{-i\phi}\,, \label{eq: rotated spinor products}
\end{equation}
one thus evaluates using the amplitude (\ref{eq: Parke-Taylor formula_trace}):
\begin{equation}
\label{eq: M++++}
\l 1_-^a2_-^b| \mathcal{M}^{(0)}|1_-'^c2_-'^d\r \delta^{cd} = 2g^2\CA \delta^{ab} \frac{1}{\cos^2\theta\sin^2\theta}\,.
\end{equation}
For $+-$ pairs, one simply inserts either $\cos^4\theta$ or $\sin^4\theta \e^{\pm4i\phi}$ into the numerator, respectively, depending on whether $1$ and $1'$ have the
same or opposite helicity; in the latter case, the sign of the phase is given by the helicity of $1'$.

\subsection{One-loop \texorpdfstring{$\beta$}{beta}-function}
\label{sec: YM eleven thirds}

\begin{figure}
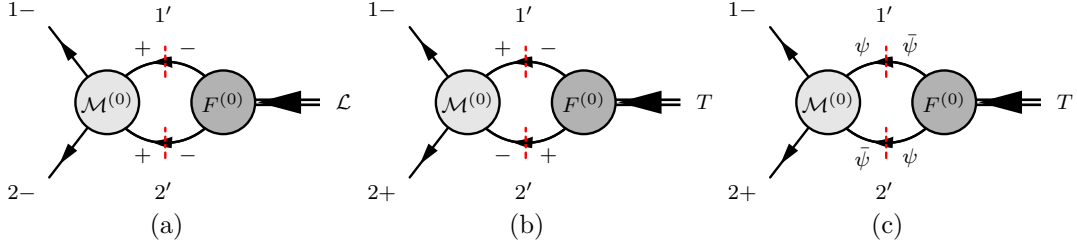

  \centering
 \begin{subfigure}[c]{0.3\textwidth}
 \centering
$
\settoheight{\eqoff}{$\times$}%
\setlength{\eqoff}{0.5\eqoff}%
\addtolength{\eqoff}{-12.0\unitlength}%
\raisebox{\eqoff}{%
\fmfframe(2,2)(2,2){%
\begin{fmfchar*}(40,20)
% outgoing
\fmfright{vq}
\fmfleft{vp2,vp1}
% internal lines left
\fmf{dbl_plain_arrow,tension=1.2}{vq,v1}
% internal lines middle
\fmf{plain_arrow,tension=0.5,left=0.7,label=$\scriptstyle 2'\,,$,l.d=15}{v1,v2}
\fmf{plain_arrow,tension=0.5,right=0.7,label=$\scriptstyle 1'\,,$,l.d=15}{v1,v2}
\fmf{phantom_smallcut,left=0.7,tension=0}{v1,v2}
\fmf{phantom_smallcut,right=0.7,tension=0}{v1,v2}
% internal lines right
\fmf{plain_arrow}{v2,vp1}
\fmf{plain_arrow}{v2,vp2}
% blobs
\fmfv{decor.shape=circle,decor.filled=30,decor.size=24,label=$\scriptstyle F^{(0)}$,label.dist=0}{v1}
\fmfv{decor.shape=circle,decor.filled=10,decor.size=24,label=$\scriptstyle \mathcal{M}^{(0)}$,label.dist=0}{v2}
\fmffreeze
 % labels  
 %%% Workaround for MetaPost 1.504 bug 
 \fmfcmd{pair vertq, vertpone, vertptwo, vertpthree, vertpL, vertone, verttwo; vertone = vloc(__v1); verttwo = vloc(__v2); vertq = vloc(__vq); vertpone = vloc(__vp1); vertptwo = vloc(__vp2);}
 % labels external states
 \fmfiv{label=$\scriptstyle \mathcal{L}$}{vertq}
 \fmfiv{label=$\scriptstyle 1-$}{vertpone}
 \fmfiv{label=$\scriptstyle 2-$}{vertptwo}
 \fmfiv{label=$\scriptstyle -$,l.d=20,l.a=-120}{vertone}
 \fmfiv{label=$\scriptstyle -$,l.d=20,l.a=+120}{vertone}
 \fmfiv{label=$\scriptstyle +$,l.d=20,l.a=-60}{verttwo}
 \fmfiv{label=$\scriptstyle +$,l.d=20,l.a=+60}{verttwo}
\end{fmfchar*}%
}}%
$   \qquad
\caption{\textcolor{white}{.}}
\label{subfig: L}
\end{subfigure}
\begin{subfigure}[c]{0.3\textwidth}
 \centering 
$
\settoheight{\eqoff}{$\times$}%
\setlength{\eqoff}{0.5\eqoff}%
\addtolength{\eqoff}{-12.0\unitlength}%
\raisebox{\eqoff}{%
\fmfframe(2,2)(2,2){%
\begin{fmfchar*}(40,20)
% outgoing
\fmfright{vq}
\fmfleft{vp2,vp1}
% internal lines left
\fmf{dbl_plain_arrow,tension=1.2}{vq,v1}
% internal lines middle
\fmf{plain_arrow,tension=0.5,left=0.7,label=$\scriptstyle 2'\,,$,l.d=15}{v1,v2}
\fmf{plain_arrow,tension=0.5,right=0.7,label=$\scriptstyle 1'\,,$,l.d=15}{v1,v2}
\fmf{phantom_smallcut,left=0.7,tension=0}{v1,v2}
\fmf{phantom_smallcut,right=0.7,tension=0}{v1,v2}
% internal lines right
\fmf{plain_arrow}{v2,vp1}
\fmf{plain_arrow}{v2,vp2}
% blobs
\fmfv{decor.shape=circle,decor.filled=30,decor.size=24,label=$\scriptstyle F^{(0)}$,label.dist=0}{v1}
\fmfv{decor.shape=circle,decor.filled=10,decor.size=24,label=$\scriptstyle \mathcal{M}^{(0)}$,label.dist=0}{v2}
\fmffreeze
 % labels  
 %%% Workaround for MetaPost 1.504 bug 
 \fmfcmd{pair vertq, vertpone, vertptwo, vertpthree, vertpL, vertone, verttwo; vertone = vloc(__v1); verttwo = vloc(__v2); vertq = vloc(__vq); vertpone = vloc(__vp1); vertptwo = vloc(__vp2);}
 % labels external states
 \fmfiv{label=$\scriptstyle T$}{vertq}
 \fmfiv{label=$\scriptstyle 1-$}{vertpone}
 \fmfiv{label=$\scriptstyle 2+$}{vertptwo}
 \fmfiv{label=$\scriptstyle +$,l.d=20,l.a=-120}{vertone}
 \fmfiv{label=$\scriptstyle -$,l.d=20,l.a=+120}{vertone}
 \fmfiv{label=$\scriptstyle -$,l.d=20,l.a=-60}{verttwo}
 \fmfiv{label=$\scriptstyle +$,l.d=20,l.a=+60}{verttwo}
\end{fmfchar*}%
}}%
$   \qquad
\caption{\textcolor{white}{.}}
\label{subfig: T gluons}
 \end{subfigure}
 \begin{subfigure}[c]{0.3\textwidth}
  \centering
$
\settoheight{\eqoff}{$\times$}%
\setlength{\eqoff}{0.5\eqoff}%
\addtolength{\eqoff}{-12.0\unitlength}%
\raisebox{\eqoff}{%
\fmfframe(2,2)(2,2){%
\begin{fmfchar*}(40,20)
% outgoing
\fmfright{vq}
\fmfleft{vp2,vp1}
% internal lines left
\fmf{dbl_plain_arrow,tension=1.2}{vq,v1}
% internal lines middle
\fmf{plain_arrow,tension=0.5,left=0.7,label=$\scriptstyle 2'\,,$,l.d=15}{v1,v2}
\fmf{plain_arrow,tension=0.5,right=0.7,label=$\scriptstyle 1'\,,$,l.d=15}{v1,v2}
\fmf{phantom_smallcut,left=0.7,tension=0}{v1,v2}
\fmf{phantom_smallcut,right=0.7,tension=0}{v1,v2}
% internal lines right
\fmf{plain_arrow}{v2,vp1}
\fmf{plain_arrow}{v2,vp2}
% blobs
\fmfv{decor.shape=circle,decor.filled=30,decor.size=24,label=$\scriptstyle F^{(0)}$,label.dist=0}{v1}
\fmfv{decor.shape=circle,decor.filled=10,decor.size=24,label=$\scriptstyle \mathcal{M}^{(0)}$,label.dist=0}{v2}
\fmffreeze
 % labels  
 %%% Workaround for MetaPost 1.504 bug 
 \fmfcmd{pair vertq, vertpone, vertptwo, vertpthree, vertpL, vertone, verttwo; vertone = vloc(__v1); verttwo = vloc(__v2); vertq = vloc(__vq); vertpone = vloc(__vp1); vertptwo = vloc(__vp2);}
 % labels external states
 \fmfiv{label=$\scriptstyle T$}{vertq}
 \fmfiv{label=$\scriptstyle 1-$}{vertpone}
 \fmfiv{label=$\scriptstyle 2+$}{vertptwo}
 \fmfiv{label=$\scriptstyle \fermp$,l.d=20,l.a=-120}{vertone}
 \fmfiv{label=$\scriptstyle \fermm$,l.d=20,l.a=+120}{vertone}
 \fmfiv{label=$\scriptstyle \fermm$,l.d=20,l.a=-60}{verttwo}
 \fmfiv{label=$\scriptstyle \fermp$,l.d=20,l.a=+60}{verttwo}
\end{fmfchar*}%
}}%
$ \qquad 
\caption{\textcolor{white}{.}}
\label{subfig: T fermions}
 \end{subfigure}
\caption{Different contributions to the anomalous dimension $\gamma_\mathcal{L}$ of the Lagrangian density and thus the Yang-Mills $\beta$-function.
This requires form factors for both the Lagrangian density (a) and stress tensor (b), with matter fields (c) contributing only to the latter.}
\label{fig: one-loop double cut}
\end{figure}

The Yang-Mills $\beta$-function is now given, according to the infrared-safe ratio in eq.~(\ref{eq: IR safe ratio}),
by acting with the above tree amplitude on the form factors
for the Lagrangian density $\mathcal{L}\equiv -G^a_{\mu\nu}G^{\mu\nu\,a}/(4g^2)$ and the stress tensor $T^{\alpha\beta,\dot\alpha\dot\beta}$.

At tree level, for each of these form factors, there is a unique polynomial in spinors that one can write down with the correct dimension, Lorentz indices,
and little-group phases:
\begin{equation}
\begin{aligned}
 \l 1^a_-2^b_-| \mathcal{L} |0\r &=\tfrac12\delta^{ab} \l12\r^2\,,
\\
 \l 1^a_- 2^b_+| T^{\alpha\beta,\dot\alpha\dot\beta}|0\r &= 2\delta^{ab}\lambda_1^\alpha\lambda_1^\beta \tlambda_2^{\dot\alpha}\tlambda_2^{\dot\beta}\,.
\end{aligned}
\end{equation}
The overall normalizations are physically meaningful and will be discussed shortly for the latter case, but they play no role for the present discussion.

To evaluate the imaginary part of the corresponding one-loop form factors,
we substitute the tree amplitude (\ref{eq: M++++}) into the phase-space integral in eq.~(\ref{eq: phase space})
as depicted in fig.~\ref{fig: one-loop double cut}\subref{subfig: L},\subref{subfig: T gluons}:
\begin{subequations}
\begin{alignat}{1}
 \l 1^a_-2^b_-|\MM \otimes \mathcal{L}|0\r^{(0)} &= \frac{2g^2\CA}{16\pi} \int \frac{\de\Omega}{4\pi}\frac{1}{\cos^2\theta\sin^2\theta} \left(\tfrac12\delta^{ab}\l1'2'\r^2\right)\,,
 \label{eq: Matrix element L}
\\
  \l 1^a_-2^b_+|\MM\otimes T^{\alpha\beta,\dot\alpha\dot\beta}|0\r^{(0)}
  &= \frac{2g^2\CA}{16\pi} \int \frac{\de\Omega}{4\pi}
  \frac{
  1}
  {\cos^2\theta\sin^2\theta}
  \Big(2\delta^{ab}\lambda'{}_1^\alpha\lambda'{}_1^\beta\tlambda'{}_2^\dalpha\tlambda'{}_2^\dbeta\,\cos^4\theta\\[-0.2\baselineskip]\nonumber
  &\hspace{0.32\textwidth}+ 2\delta^{ab}\tlambda'{}_1^{\dot\alpha}\tlambda'{}_1^{\dot\beta}\lambda'{}_2^\alpha\lambda'{}_2^\beta\,\sin^4\theta \e^{4i\phi}
  \Big) 
  \,.
\end{alignat}\end{subequations}
Note that the tree form factors are evaluated with the rotated spinors (\ref{rotation}) parametrizing the two intermediate states in the cut.
The two terms in the last line come from the two possible intermediate helicities, of which only one is shown in fig.~\ref{subfig: T gluons}.
A priori, they look quite complicated; expanding out the first gives
\begin{equation}
\begin{aligned}
\lambda'{}_1^\alpha\lambda'{}_1^\beta\tlambda'{}_2^\dalpha\tlambda'{}_2^\dbeta &=
(\lambda_1\cos\theta-\lambda_2\sin\theta \e^{i\phi})^\alpha
(\lambda_1\cos\theta-\lambda_2\sin\theta \e^{i\phi})^\beta
\\ &\hspace{5mm}\times (\tlambda_2\cos\theta+\tlambda_1\sin\theta \e^{i\phi})^{\dot\alpha} (\tlambda_2\cos\theta+\tlambda_1\sin\theta \e^{i\phi})^{\dot\beta}\,.
\end{aligned}
\end{equation}
However, the key is that ultimately the spinor structure is fixed by little-group weights, which are enforced by the azimuthal angle integration.
Indeed, we see that all terms with non-vanishing phases are killed by the $\phi$ integration!
Dropping these, the integral becomes simply proportional to the tree form factor,
as anticipated below eq.~(\ref{eq: gammaO practical}).
Hence, the ratio does not depend on the spinor indices and
\begin{equation} \label{eq: stress tensor YM}
 \frac{\l 1^a_-2^b_+|\MM \otimes T^{\alpha\beta,\dot\alpha\dot\beta}|0\r^{(0)}}{\l 1^a_-2^b_+|T^{\alpha\beta,\dot\alpha\dot\beta}|0\r^{(0)}}
 =  \frac{2g^2C_A}{16\pi} \int_0^{\frac{\pi}{2}} 2\sin\theta\cos\theta\de\theta \frac{\cos^8\theta+\sin^8\theta}{\cos^2\theta\sin^2\theta}\,.
\end{equation}
Now we can observe that the divergences in the collinear limits $\theta\to 0,\pi/2$ cancel precisely against those in the Lagrangian density
in (\ref{eq: gammaO practical}), yielding, as anticipated, a convergent integral:
\begin{align}
 \gamma^{(1)}_{\mathcal{L}} &\equiv -\frac{1}{\pi} \left(
 \frac{\l 1^a_-2^b_-|\MM \otimes \mathcal{L}|0\r^{(0)}}{\l 1^a_-2^b_-|\mathcal{L}|0\r^{(0)}} -
 \frac{\l 1^a_-2^b_+|\MM\otimes T^{\alpha\beta,\dot\alpha\dot\beta}|0\r^{(0)}}{\l 1^a_-2^b_+|T^{\alpha\beta,\dot\alpha\dot\beta}|0\r^{(0)}}
 \right)
\nl &= -\frac{2g^2\CA}{16\pi^2} \int_0^{\frac{\pi}{2}} 2\sin\theta\cos\theta\de\theta \left(\frac{1}{\cos^2\theta\sin^2\theta}-\frac{\cos^8\theta+\sin^8\theta}{\cos^2\theta\sin^2\theta}\right)
\nl &= -\frac{2g^2}{16\pi^2} \times \frac{11\CA}{3}\,. \label{eq: beta function YM}
\end{align}
Using the relation between the running of the Yang-Mills Lagrangian and the $\beta$-function
quoted earlier, eq.~(\ref{eq:gamma vs beta function}), we have therefore obtained the one-loop $\beta$-function:
\begin{equation}
\beta(g^2)= -\frac{2g^4}{16\pi^2}\times \frac{11\CA}{3}\,, \qquad \mbox{where}\qquad \beta(g^2)\equiv \mu\partial_\mu g^2(\mu) \,.
\end{equation}
This is in perfect agreement with the famous result, including, of course, the sign!

This example confirms that one-loop anomalous dimensions can be obtained as suitable differences between
eigenvalues of the tree-level $S$-matrix, or more precisely, of minus the phase of the $S$-matrix divided by $\pi$.
In the case above, the scattering phase is positive ($\mathcal{M}>0$ in eq.~(\ref{eq: M++++})), which is attributed to the attractive nature
of the interaction between opposite color charges (the scattering phase represents, roughly, minus the interaction energy).
This is the reason in this framework for the famous negative sign of the $\beta$-function. 
More precisely, the reason is that the attraction is felt more strongly in the $s$-wave state (Lagrangian density) than in the $d$-wave state (stress tensor).

\subsection{Matter-field contributions} \label{ssec: matter}

It is instructive to see how the method works in the presence of fermions and scalars coupled to the Yang-Mills field.
Naively, since the Yang-Mills part of the Lagrangian density has no tree-level coupling to matter, one might worry that its anomalous dimension would be insensitive to these.
However, the infrared structure of the theory is modified and this is detected by the stress tensor in the denominator of the IR-safe ratio (\ref{eq: IR safe ratio}).
In QED, this would be the only contribution.

To find out how the stress tensor couples to fermions and scalars, one could
construct the stress tensor following the Noether procedure and apply standard Feynman rules.
We use a shortcut exploiting the symmetries of the problem.
The overall normalization (at least, relative to the gluon contribution)
will be important. It is fixed physically by requiring that the expectation value of the stress tensor in a state returns its momentum \cite{Peskin:1995ev}:
\begin{equation}
 \l 1_\Phi | T^{\alpha\beta,\dot\alpha\dot\beta}|1_\Phi\r = 2p_1^{\alpha\dot\alpha}p_1^{\beta\dot\beta} = \l 1_\Phi\bar{1}_{\bar{\Phi}}| T^{\alpha\beta,\dot\alpha\dot\beta}|0\r\,,
\end{equation}
where in the second step we used crossing symmetry. Thus, the forward limit $p_2\to -p_1$ of the form factor is fixed.
For fermions there is an analogous equation, but one needs to be mindful of the
sign in the crossing relation (\ref{eq: crossing}) for each $\fermm$ in the initial state, so the condition is
\begin{equation}
 \l 1_{\fermm} | T^{\alpha\beta,\dot\alpha\dot\beta}|1_{\fermm}\r = 2p_1^{\alpha\dot\alpha}p_1^{\beta\dot\beta} = -\l 1_{\fermm} \bar{1}_{\fermp}| T^{\alpha\beta,\dot\alpha\dot\beta}|0\r\,.
\end{equation}
The other constraint is that the stress tensor is conserved: it must be orthogonal to $(p_1+p_2)$.
For scalars, as is well-known, this leaves an ambiguity which can be removed by imposing tracelessness (equivalent to symmetry in the spinor indices).
For both scalars and fermions, there is then a unique polynomial satisfying these constraints and little-group scaling:
\begin{equation}
\label{eq: form factor of energy momentum YM with fermions}
\begin{aligned}
 \l 1_{\bar\Phi} 2_{\Phi}|T^{\alpha\beta,\dot\alpha\dot\beta}|0\r &= \tfrac13\big(
 p_1^{\alpha\dalpha}p_1^{\beta\dbeta}+p_2^{\alpha\dalpha}p_2^{\beta\dbeta}
-p_1^{\alpha\dalpha}p_2^{\beta\dbeta}-p_1^{\beta\dalpha}p_2^{\alpha\dbeta}-p_1^{\alpha\dbeta}p_2^{\beta\dalpha}-p_1^{\beta\dbeta}p_2^{\alpha\dalpha}\big)\,,\\
  \l 1_{\fermm} 2_{\fermp}|T^{\alpha\beta,\dot\alpha\dot\beta}|0\r &=
 \tfrac{1}{2}
 \big(\lambda_1^\alpha\lambda_1^\beta \tlambda_1^{\dot\alpha}\tlambda_2^{\dot\beta}+
 \lambda_1^\alpha\lambda_1^\beta \tlambda_1^{\dot\beta}\tlambda_2^{\dot\alpha}-
 \lambda_1^\alpha\lambda_2^\beta \tlambda_2^{\dot\alpha}\tlambda_2^{\dot\beta}-
 \lambda_1^\beta\lambda_2^\alpha \tlambda_2^{\dot\alpha}\tlambda_2^{\dot\beta}\big)\,.
\end{aligned}\end{equation}
In accordance with eq.~(\ref{eq: gammaO practical}), we now convolute these form factors with annihilation amplitudes into two gluons,
as illustrated in fig.~\ref{fig: one-loop double cut}\subref{subfig: T fermions} for fermions.
The relevant tree amplitudes are all concisely encoded in an $\mathcal{N}=4$ supersymmetric expression
using Nair's $\mathcal{N}=4$ on-shell superspace \cite{Nair:1988bq}, which generalizes the amplitude (\ref{eq: Parke-Taylor formula_trace}) to
\begin{equation} \label{eq: Nair superamplitude}
\mathcal{M}^{abcd}_{1234}\delta^{cd}=-2g^2\CA\delta^{ab}
\frac{\delta^8(\sum_{i=1}^{4}\lambda_i \tildeeta_i)}{\l13\r\l32\r\l24\r\l 41\r}\equiv 
-2g^2\CA\delta^{ab}\frac{\prod_{A=1}^4 \sum_{1\leq i<j\leq4}\l ij\r \tildeeta_i\tildeeta_j}{\l13\r\l32\r\l24\r\l 41\r}\,.
\end{equation}
To insert a negative-helicity gluon, fermion, scalar or positive-helicity fermion on site $j$, one extracts, respectively,
four, three, two or one powers of $\tildeeta_j$, giving the required amplitudes:
\begin{equation}
\label{eq: annihilation amplitude}
\begin{aligned}
 \l 1^a_-2^b_+|\mathcal{M}|1_{\Phi}'2_{\bar\Phi}'\r &=
  -2g^2n_sT_s \delta^{ab} \frac{\l11'\r^2\l12'\r^2}{\l 11'\r\l12'\r\l21'\r\l22'\r}=2g^2n_sT_s \delta^{ab} \frac{\cos^2\theta\sin^2\theta \e^{2i\phi}}{\cos^2\theta\sin^2\theta}\,,
 \\
 \l 1^a_-2^b_+|\mathcal{M}|1_{\fermm}'2_{\fermp}'\r &=
  -2g^2(2n_f)T_f \delta^{ab} \frac{\l11'\r\l12'\r^3}{\l 11'\r\l12'\r\l21'\r\l22'\r}=-4g^2n_fT_f \delta^{ab} \frac{\cos^3\theta\sin\theta \e^{i\phi}}{\cos^2\theta\sin^2\theta}\,,
 \\
 \l 1^a_-2^b_+|\mathcal{M}|1_{\fermp}'2_{\fermm}'\r &=
  -2g^2(2n_f)T_f \delta^{ab} \frac{\l11'\r^3\l12'\r}{\l 11'\r\l12'\r\l21'\r\l22'\r}=-4g^2n_fT_f \delta^{ab} \frac{\cos\theta\sin^3\theta \e^{3i\phi}}{\cos^2\theta\sin^2\theta}\,,
\end{aligned}
\end{equation}
and $\l 1^a_-2^b_+|\mathcal{M}|1_{\bar\Phi}'2_{\Phi}'\r=\l 1^a_-2^b_+|\mathcal{M}|1_{\Phi}'2_{\bar\Phi}'\r$.
Here, anticipating the contraction with the stress tensor, we have re-weighted the color-adjoint amplitude (\ref{eq: Nair superamplitude}) in accordance
to $n_s$ complex scalars and $n_f$ Dirac fermions (and thus $(2n_f)$ Weyl fermions) in representations where $\Tr[T^aT^b]=T_{s,f}\delta^{ab}$, with $T_F=\tfrac12$ in the fundamental representation.
The final step is to integrate this over phase space, weighted by the tree form factors in eqs.~(\ref{eq: form factor of energy momentum YM with fermions})
evaluated with the rotated spinors (\ref{rotation}).
Again, most terms drop out upon azimuthal integration, leaving, as expected, a result proportional to the tree form factor:
\begin{align}\label{eq: stress tensor QCD}
 \frac{\l 1^a_-2^b_+|\MM \otimes T^{\alpha\beta,\dot\alpha\dot\beta}|0\r^{(0)}}{
 \l 1^a_-2^b_+|T^{\alpha\beta,\dot\alpha\dot\beta}|0\r^{(0)}} &\simeq
\frac{2g^2}{16\pi} \int_0^{\frac{\pi}{2}} 
\frac{2\sin\theta\cos\theta \de\theta}{\cos^2\theta\sin^2\theta}\Big[\CA(\cos^8\theta+\sin^8\theta) 
\\ & \hspace{10mm} +2 n_fT_f(\cos^6\theta\sin^2\theta +\sin^6\theta\cos^2\theta)+ 2n_sT_s \cos^4\theta\sin^4\theta\Big]\,. \nonumber
\end{align}
As a simple check, one can plug in the matter content of $\mathcal{N}=4$ SYM
(two adjoint Dirac fermions and three complex scalars: $n_f{=}2$, $n_s{=}3$, $T_f{=}T_s{=}\CA$),
and see that the bracket reduces to $\CA(\cos^2\theta+\sin^2\theta)^4=\CA$. This reproduces
the integrand for the Lagrangian density in eq.~(\ref{eq: Matrix element L}), as required by supersymmetry
since the stress tensor and Lagrangian density are in the same supermultiplet. The vanishing of the $\beta$-function in $\mathcal{N}=4$ is thus automatic in this formalism
and can be used as a simple check on the algebra.
For other theories, replacing the subtraction in eq.~(\ref{eq: beta function YM}) by (\ref{eq: stress tensor QCD}) and integrating, we reproduce
the well-known one-loop result for general matter content:
\ba
\beta(g^2)&=& -\frac{2g^4}{16\pi^2}b_0\,,\qquad b_0\equiv \frac{11}{3}C_A -\frac{4}{3}n_f T_f -\frac{1}{3} n_s T_s\,.
\ea

\def\Ffermm{\fermm_F}
\def\Ffermp{\fermp_F}
\def\Fbfermm{\fermm_{\bar F}}
\def\Fbfermp{\fermp_{\bar F}}
In a theory with fermion masses like QCD, the running of mass parameters is also interesting.
At energies much higher than the masses (the situation where ``running" is meaningful), we expect this question
to be answerable within the massless theory.  Writing a Dirac fermion as a combination
of positive- and negative-helicity fundamental Weyl fermions $\Psi=(\Ffermp,\Ffermm)$ and complex conjugate $\bar\Psi=(\Fbfermm,\Fbfermp)$,
the minimal form factor for the mass operator $\bar{\Psi}\Psi=\bar{\Psi}_a\Psi^a=(\Fbfermm{}_a\Ffermm^a+\Fbfermp{}_a\Ffermp^a)$
is $\l 1_{\Ffermm}2_{\Fbfermm}|\bar{\Psi}\Psi|0\r=\l12\r$.
The required scattering amplitudes between fundamental and antifundamental fermions, for same and opposite helicity respectively, are then
\be
\l  1_{\Ffermm}2_{\Fbfermm}|\MM| 1'_{\Ffermm}2'_{\Fbfermm}\r =
\frac{2g^2\CF}{\cos^2\theta\sin^2\theta}\,,
\qquad
\l  1_{\Ffermm}2_{\Fbfermp}|\MM| 1'_{\Ffermm}2'_{\Fbfermp}\r=\frac{2g^2\CF\cos^4\theta}{\cos^2\theta\sin^2\theta}\,,
\ee
where the fundamental Casimir is $\CF=\frac{N_c^2-1}{2N_c}$ for gauge group SU(N$_c$). The positive signs again reflect the attractive gauge interaction.
We also need the pair production amplitude $\l 1_{\Ffermm}2_{\Fbfermp}|\MM|1'_-2'_+\r$, equal to minus the complex conjugate of (\ref{eq: annihilation amplitude}).
Armed with these and the above stress-tensor form factors for gluons and fermions, we compute
\begin{align}
 \gamma^{(1)}_{\bar\Psi\Psi} &\equiv -\frac{1}{\pi} \left(
 \frac{\l  1_{\Ffermm}2_{\Fbfermm}|\MM \otimes \bar\Psi\Psi|0\r^{(0)}}{\l  1_{\Ffermm}2_{\Fbfermm}| \bar\Psi\Psi|0\r^{(0)}} -
 \frac{\l 1_{\Ffermm}2_{\Fbfermp}|\MM\otimes T^{\alpha\beta,\dot\alpha\dot\beta}|0\r^{(0)}}{\l 1_{\Ffermm}2_{\Fbfermp}|T^{\alpha\beta,\dot\alpha\dot\beta}|0\r^{(0)}}
 \right)
\nl &= -\frac{2g^2\CF}{16\pi^2} \int_0^{\frac{\pi}{2}} 2\sin\theta\cos\theta\de\theta \left(\frac{1}{\cos^2\theta\sin^2\theta}-\frac{\cos^6\theta+\sin^6\theta}{\cos^2\theta\sin^2\theta}\right)
= -\frac{6g^2\CF}{16\pi^2}\,. \label{eq: fermion mass running}
\end{align}
Again, this is in agreement with the standard result, confirming that running-mass effects at short distances can be computed using unitarity with massless states.

\subsubsection{Comments on masses}
\label{eq: comment on masses}

Our discussions so far have been restricted to the $S$-matrix of strictly massless particles -- the dilatation operator $D$ is only defined on the massless $S$-matrix!
We believe that this is not a significant restriction. Rather, we believe it is entirely consistent with conventional applications of the renormalization group, where
a particle is either regarded as heavy and integrated out, or as light, in which case its mass is neglected.
These two effective descriptions are connected by so-called matching regions where the masses are important,
but which do not produce the kind of large logarithms that the renormalization group resums and which are the focus of this paper.
The running of relevant operators such as QCD masses can be correctly calculated within the massless theory in the high-energy regime,
as we have just explicitly verified.

With massive particles, one can get in addition momentum-independent logarithms.
For example, a massive tadpole integral\footnote{%
Such logarithms can appear from any integral with an explicit mass, not necessarily of tadpole topology.}  gives
\be
 \int \frac{\mu^{2\peps}\de^{4{-}2\peps}l}{i (2\pi)^{4{-}2\peps}} \frac{1}{l^2-m^2}= \frac{m^2}{16\pi^2}\left(\frac{1}{\peps}+\log\frac{\mu^2}{m^2}+\ldots\right)\,. \label{eq: massless log}
\ee
It is common in textbook presentations of the renormalization group to focus on ultraviolet divergences and therefore include such logarithms when computing $\beta$-functions.
Yet these logarithms lack an imaginary part and so they cannot be detected by unitarity.
Does this mean that the unitarity method is incomplete?  We believe no, because the renormalization group can answer two distinct questions.
The first type of question regards the running of \emph{bare} couplings as a function of the short-distance cutoff.  This is clearly of importance to lattice practitioners, for example.
The above momentum-independent logarithms are then clearly relevant (and possibly also power divergences as well as details of the short-distance dynamics).
The second type of question regards the optimal coupling to use to minimize large logarithms, for example in the perturbative calculation of a cross-section at a given energy scale.
This is the typical question of interest to collider physicists.
Momentum-independent logarithms are then clearly \emph{not} relevant: once the bare couplings have been tuned to cancel $\log(\mu)$ for one observable, the same tuning removes it from any physical observable.
Our conclusion is that unitarity, by throwing away the logarithms (\ref{eq: massless log}),\footnote{%
In massless contexts with evanescent operators, a similar distinction between physically observable logarithms versus bare ultraviolet divergences (poles in dimensional regularization) is also important \cite{Bern:2015xsa}.} correctly answers the second type of question.

\subsection{Twist-two operators and partial-wave amplitudes}
\def\ttheta{\theta}

A pleasant feature of the unitarity approach is that the $S$-matrix for just a few basic processes controls the anomalous dimension of essentially any operator.
Let us here discuss those operators which can be accessed using a color-singlet pair of partons, as considered so far.

Let us ignore spin for a moment and consider for simplicity two-particle form factors for a complex scalar.
Tree form factors are polynomial in $p_1^\mu,p_2^\mu$.
Factors of $(p_1{+}p_2)^\mu$ represent uninteresting total derivatives; these can be projected out by considering the forward case $p_2=-p_1$.
As $p_1^2=0$, only traceless tensors then survive.
Thus, the interesting polynomials of order $j$ represent operators of spin $j$ and dimension $j+2$ (the plus two is because any external on-shell parton carries dimension 1).
These are the form factors of twist-two operators:
\begin{equation}
\l 1_\Phi \bar{1}_{\bar\Phi}|\OO_m| 0\r = p_1^{\mu_1}\cdots p_1^{\mu_m} \quad\Leftrightarrow\quad
\OO_m = i^n \,\bar\Phi \partial^{\mu_1}\cdots \partial^{\mu_m}\Phi\,.
\end{equation}
Let us act on these polynomial with the tree-level $S$-matrix.
Note that, even though this action is originally derived assuming that all particles carry positive energy, in the spinor
parametrization (\ref{rotation}) the phase-space integrals can be seamlessly continued to the forward case $p_2=-p_1$.
The rotated form factor in this parametrization is then a multiple of itself, since
\begin{equation}
 p_1'^{\alpha\dalpha} \equiv \lambda_1'^\alpha\tilde{\lambda}_1'^{\dalpha} =\lambda_1^\alpha\tilde{\lambda}_1^{\dalpha}(\cos\theta-\sin\theta \e^{i\phi})(\cos\theta+\sin\theta \e^{-i\phi})
= p_1^{\alpha\dalpha}(\cos(2\theta)-i\sin(2\theta)\sin\phi).
\end{equation}
Using that the $S$-matrix for scalars does not depend on the azimuthal part of the scattering angle,
the latter can be integrated out immediately.
Temporarily rescaling the scattering angle $2\theta\to \ttheta$, we find that this
produces Legendre polynomials:
\begin{equation}
\int_0^{2\pi}\frac{\de\phi}{2\pi}\big(\!\cos\ttheta-i\sin\ttheta\sin\phi\big)^m=P_m(\cos\theta).
\end{equation}
For two complex scalars, the basic unitarity relation (\ref{eq: gamma UV plus IR}) thus becomes
\begin{equation}
 \gamma^{(1)}_{\OO_m} - \gamma^{(1)}_\IR = -\frac{1}{16\pi^2}  \int_0^{\pi} \frac{\sin\ttheta \de\ttheta}{2} \MM^{(0)}(\cos\ttheta) P_m(\cos\theta) \equiv -\frac{1}{\pi} a_m^{(0)}\,.
\end{equation}
We recognize $a_m$ as the partial-wave amplitude with angular momentum $m$, which at leading order can be identified
with the phase of the $S$-matrix, normalized as $S_m=1+i\mathcal{M}_m = \e^{ia_m}$.
Thus, anomalous dimension are indeed minus the phase of the $S$-matrix, divided by $\pi$, as expected from eq.~\eqref{cute_eigenvalue_equation},
and two-particle states with definite angular momentum map to twist-two operators.

Let us apply this to a few examples.  First consider twist-two operators with two identical complex scalars $Z$ in $\mathcal{N}=4$ SYM:
$\OO_m=\Tr[Z\partial^+_mZ]$, where $m$ is even.  Because it is in the same multiplet, the tree amplitude is the same as in eq.~(\ref{eq: Matrix element L}). Using the stress tensor
to subtract the infrared divergence via eq.~(\ref{eq: stress tensor QCD}), the formula becomes
\begin{equation}
 \gamma^{(1)}_{\OO_m} = -\frac{2g^2N_c}{16\pi^2} \int_0^{\pi} \frac{2\de\ttheta}{\sin\ttheta} \left(P_m(\cos\theta)-P_m(1)\right) = \frac{g^2N_c}{16\pi^2}\times  8S_1(m)\,,
\end{equation}
where $S_1(m)=\sum_{i=1}^m \tfrac{1}{i}$ denotes the harmonic sum. This is precisely the know result \cite{Kotikov:2000pm}.

In pure Yang-Mills, the similar partial-wave analysis requires partial waves for particles with spin.
These are more complicated than Legendre polynomials but the spinor parametrization provides a straightforward way to proceed.
Let us first record a formula for the evolution of an arbitrary operator which can decay to two particles at tree level,
which follows by combining the unitarity relation (\ref{eq: gammaO practical}), the matrix element (\ref{eq: M++++})
and the stress-tensor eigenvalue (\ref{eq: stress tensor QCD}):
\begin{equation}
\begin{aligned}
 \gamma_\OO^{(1)} \l1_-2_-| \OO|0\r^{(0)} &= \frac{g^2C_A}{16\pi^2} \int_0^{2\pi} \frac{\de\phi}{2\pi}\int_0^{\frac{\pi}{2}}\frac{4\de\theta}{\cos\theta\sin\theta}
 \left[\begin{array}{r}\big(\cos^8\theta+\sin^8\theta\big)\l1_-2_-| \OO|0\r^{(0)}\\-\l1_-'2_-'| \OO|0\r^{(0)}\end{array}\right],
\\
 \gamma_\OO^{(1)} \l1_-2_+| \OO|0\r^{(0)} &= \frac{g^2C_A}{16\pi^2} \int_0^{2\pi} \frac{\de\phi}{2\pi}\int_0^{\frac{\pi}{2}}\frac{4\de\theta}{\cos\theta\sin\theta}
\left[\begin{array}{r}
\big(\cos^8\theta+\sin^8\theta\big)\l1_-2_+| \OO|0\r^{(0)}\\
 -\cos^4\theta\l1_-'2_+'| \OO|0\r^{(0)}\\
 -\sin^4\theta \e^{4i\phi}\l1_+'2_-'| \OO|0\r^{(0)}\end{array}\right].
\end{aligned}\label{mainresult}
\end{equation}
In Yang-Mills theory, the leading-twist operators can be either in the vector-like Lorentz representation
$(\tfrac{m}{2},\tfrac{m}{2})$, or in chiral representations $(\frac{m}{2}+1,\frac{m}{2}-1)$ with $m\geq 1$.
Focusing on the former, which control the energy dependence of unpolarized parton distribution functions and are associated with
the polynomials ($m\geq 2$)\footnote{%
Note that even though $\lambda_2^\alpha \tilde{\lambda}_2^{\dalpha}\simeq -\lambda_1^\alpha \tilde{\lambda}_1^{\dalpha}$
has been used to simplify the form factor in the forward limit, we have not used the stronger condition $\tilde{\lambda}_2^{\dalpha}\simeq -\tilde{\lambda}_1^{\dalpha}$
to eliminate $\tilde\lambda_2$ because the phase-space integral using eq.~(\ref{rotation}) produces additional little-group phases that do not preserve this relation.
}
\begin{equation}
 \l 1_-\bar{1}_+|\OO_{gg,m}|0\r= (\lambda_1^1\tilde{\lambda}_2^{\dot1})^{2} (\lambda_1^1\tilde{\lambda}_1^{\dot 1})^{m-2},
\end{equation}
this gives
\begin{multline}
 \gamma_{gg,m}^{(1)} =
 \frac{g^2C_A}{16\pi^2} \int_0^{2\pi} \frac{\de\phi}{2\pi}\int_0^{\frac{\pi}{2}}\frac{4\de\theta}{\cos\theta\sin\theta}
\big[
\cos^8\theta+\sin^8\theta\\ \qquad \qquad \qquad \,\,\,\,
- \cos^4\theta(\cos\theta{-}\sin\theta \e^{i\phi})^{m{+}2}(\cos\theta{+}\sin\theta \e^{-i\phi})^{m{-}2}
 \\
 -\sin^4\theta \e^{4i\phi}(\cos\theta{-}\sin\theta \e^{i\phi})^{m{-}2}(\cos\theta{+}\sin\theta \e^{-i\phi})^{m{+}2}
 \big]\,. 
%  \end{aligned}
 \label{eq: YM evolution}
\end{multline}
We have checked for several values of $m$ that this reproduces precisely the moments of the DGLAP parton evolution
equation in Yang-Mills theory,
\be
 \gamma_{gg,m} = -\int_0^1 \de x x^{m{-}1} P_{gg}(x)\,,\quad
 P_{gg}^{(1)}(x) = \frac{2g^2\CA}{16\pi^2}\left[
 2\frac{1+x^4+(1-x)^4}{x(1-x)_+} +\frac{11}{3}\delta(1{-}x)\right]\,, \label{eq: DGLAP}
\ee
as expected from the standard relation between twist-two operators and parton distribution functions \cite{Peskin:1995ev}.
Therefore, the tree-level scattering phases in Yang-Mills theory are indeed
the same as the anomalous dimensions of twist-two operators.
It would be nice to find a more direct mathematical map between eqs.~(\ref{eq: YM evolution}) and (\ref{eq: DGLAP}).

\begin{figure}[tbp]
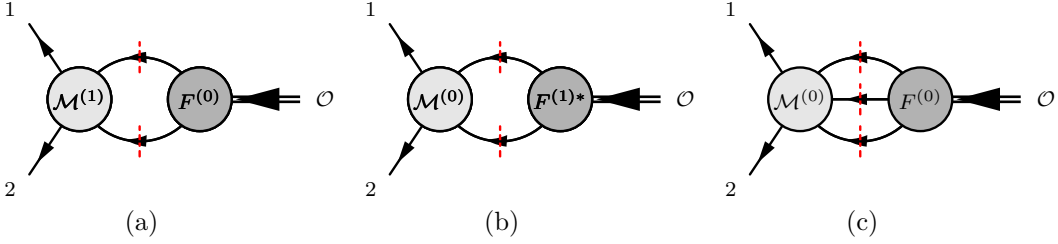
	
\centering
\begin{subfigure}[t]{0.3\textwidth}
 \centering
$
\settoheight{\eqoff}{$\times$}%
\setlength{\eqoff}{0.5\eqoff}%
\addtolength{\eqoff}{-12.0\unitlength}%
\raisebox{\eqoff}{%
\fmfframe(2,2)(2,2){%
\begin{fmfchar*}(40,20)
% outgoing
\fmfright{vq}
\fmfleft{vp2,vp1}
% internal lines left
\fmf{dbl_plain_arrow,tension=1.2}{vq,v1}
% internal lines middle
\fmf{plain_arrow_smallcut,left=0.7,tension=0.5}{v1,v2}
\fmf{plain_arrow_smallcut,right=0.7,tension=0.5}{v1,v2}
% internal lines right
\fmf{plain_arrow,tension=1.2}{v2,vp1}
\fmf{plain_arrow,tension=1.2}{v2,vp2}
% % blobs
 \fmfv{decor.shape=circle,decor.filled=30,decor.size=24,label=$\scriptstyle F^{(0)}$,label.dist=0}{v1}
 \fmfv{decor.shape=circle,decor.filled=10,decor.size=24,label=$\scriptstyle \mathcal{M}^{(1)}$,label.dist=0}{v2}
\fmffreeze
% \fmf{phantom_smallcut,left=0.7,tension=0.25}{v1,v2}
% \fmf{phantom_smallcut,right=0.7,tension=0.25}{v1,v2}
% \fmfdraw
 % labels  
 %%% Workaround for MetaPost 1.504 bug 
 \fmfcmd{pair vertq, vertpone, vertptwo, vertpthree, vertpL, vertone, verttwo; vertone = vloc(__v1); verttwo = vloc(__v2); vertq = vloc(__vq); vertpone = vloc(__vp1); vertptwo = vloc(__vp2);}
 % vertices
%
\fmfiv{decor.shape=circle,decor.filled=30,decor.size=24,label=$\scriptstyle F^{(0)}$,label.dist=0}{vertone}
\fmfiv{decor.shape=circle,decor.filled=10,decor.size=24,label=$\scriptstyle \mathcal{M}^{(1)}$,label.dist=0}{verttwo}
\fmfiv{decor.shape=circle,decor.filled=0,decor.size=19,label=$\scriptstyle \mathcal{M}^{(1)}$,label.dist=0}{verttwo}
\fmfdraw
 % labels external states
 \fmfiv{label=$\scriptstyle \mathcal{O}$}{vertq}
 \fmfiv{label=$\scriptstyle 1$}{vertpone}
 \fmfiv{label=$\scriptstyle 2$}{vertptwo}
\end{fmfchar*}%
}}%
$
\caption{\textcolor{white}{.}}
\label{fig: one two-loop double cut}
\end{subfigure}
\begin{subfigure}[t]{0.3\textwidth}
 \centering
$
\settoheight{\eqoff}{$\times$}%
\setlength{\eqoff}{0.5\eqoff}%
\addtolength{\eqoff}{-12.0\unitlength}%
\raisebox{\eqoff}{%
\fmfframe(2,2)(2,2){%
\begin{fmfchar*}(40,20)
% outgoing
\fmfright{vq}
\fmfleft{vp2,vp1}
% internal lines left
\fmf{dbl_plain_arrow,tension=1.2}{vq,v1}
% \fmf{plain_arrow,tension=0}{v1,vpL}
% \fmf{plain_arrow,tension=1.2}{v1,vp3}
% internal lines middle
\fmf{plain_arrow,left=0.7,tension=0.25}{v1,v2}
\fmf{plain_arrow,right=0.7,tension=0.25}{v1,v2}
\fmf{phantom_smallcut,left=0.7,tension=0.25}{v1,v2}
\fmf{phantom_smallcut,right=0.7,tension=0.25}{v1,v2}
% internal lines right
\fmf{plain_arrow,tension=1.2}{v2,vp1}
\fmf{plain_arrow,tension=1.2}{v2,vp2}
% blobs
\fmfv{decor.shape=circle,decor.filled=10,decor.size=24,label=$\scriptstyle \mathcal{M}^{(0)}$,label.dist=0}{v2}
\fmfv{decor.shape=circle,decor.filled=30,decor.size=24,label=$\scriptstyle F^{(1)*}$,label.dist=0}{v1}
% % \fmfv{decor.shape=circle,decor.filled=0,decor.size=20}{v1}
% \fmfv{decor.shape=circle,decor.filled=10,decor.size=20,label=$\scriptstyle \mathcal{M}$,label.dist=0}{v2}
\fmffreeze
\fmfdraw
 % labels  
 %%% Workaround for MetaPost 1.504 bug 
 \fmfcmd{pair vertq, vertpone, vertptwo, vertpthree, vertpL, vertone, verttwo; vertone = vloc(__v1); verttwo = vloc(__v2); vertq = vloc(__vq); vertpone = vloc(__vp1); vertptwo = vloc(__vp2);}
 % vertex
\fmfiv{decor.shape=circle,decor.filled=10,decor.size=24,label=$\scriptstyle \mathcal{M}^{(0)}$,label.dist=0}{verttwo}
\fmfiv{decor.shape=circle,decor.filled=30,decor.size=24,label=$\scriptstyle F^{(1)*}$,label.dist=0}{vertone}
\fmfiv{decor.shape=circle,decor.filled=0,decor.size=19,label=$\scriptstyle F^{(1)*}$,label.dist=0}{vertone}
 % labels external states
 \fmfiv{label=$\scriptstyle \mathcal{O}$}{vertq}
 \fmfiv{label=$\scriptstyle 1$}{vertpone}
 \fmfiv{label=$\scriptstyle 2$}{vertptwo}
\end{fmfchar*}%
}}%
$
\caption{\textcolor{white}{.}}
\label{fig: other two-loop double cut}
\end{subfigure}
\begin{subfigure}[t]{0.3\textwidth}
 \centering
$
\settoheight{\eqoff}{$\times$}%
\setlength{\eqoff}{0.5\eqoff}%
\addtolength{\eqoff}{-12.0\unitlength}%
\raisebox{\eqoff}{%
\fmfframe(2,2)(2,2){%
\begin{fmfchar*}(40,20)
% outgoing
\fmfright{vq}
\fmfleft{vp2,vp1}
% internal lines left
\fmf{dbl_plain_arrow,tension=1.2}{vq,v1}
% internal lines middle
\fmf{plain_arrow,left=0.7,tension=0.25}{v1,v2}
\fmf{plain_arrow,right=0.7,tension=0.25}{v1,v2}
\fmf{phantom_smallcut,left=0.7,tension=0.25}{v1,v2}
\fmf{phantom_smallcut,right=0.7,tension=0.25}{v1,v2}
\fmf{plain_arrow,tension=0,tag=1}{v1,v2}
\fmf{phantom_smallcut,tension=0}{v1,v2}
% internal lines right
\fmf{plain_arrow,tension=1.2}{v2,vp1}
\fmf{plain_arrow,tension=1.2}{v2,vp2}
% blobs
\fmfv{decor.shape=circle,decor.filled=30,decor.size=24,label=$\scriptstyle F^{(0)}$,label.dist=0}{v1}
\fmfv{decor.shape=circle,decor.filled=10,decor.size=24,label=$\scriptstyle \mathcal{M}^{(0)}$,label.dist=0}{v2}
\fmffreeze
 % labels  
 %%% Workaround for MetaPost 1.504 bug 
 \fmfcmd{pair vertq, vertpone, vertptwo, vertpthree, vertpL, vertone, verttwo; vertone = vloc(__v1); verttwo = vloc(__v2); vertq = vloc(__vq); vertpone = vloc(__vp1); vertptwo = vloc(__vp2);}
 % labels external states
 \fmfiv{label=$\scriptstyle \mathcal{O}$}{vertq}
 \fmfiv{label=$\scriptstyle 1$}{vertpone}
 \fmfiv{label=$\scriptstyle 2$}{vertptwo}
 % label middle cuts
\fmfipath{p[]}
\fmfiset{p1}{vpath1(__v1,__v2)}
% \fmfiv{label=$\scriptstyle l_2\,,$,l.d=5,l.a=-115}{point length(p1)/2 of p1}
\end{fmfchar*}%
}}%
$
\caption{\textcolor{white}{.}}
\label{fig: two-loop p1+p2 triple cut}
\end{subfigure}
\caption{Cut diagrams which contribute to the product $\mathcal{M}F^*$ at two-loop order.
}
\label{fig: two-loop cuts}
\end{figure}

At higher loops, we warn the reader that since $S$ is a matrix, its \emph{phase} (as defined from its eigenvalues) need not agree with the phase of $2{\to}2$ $S$-matrix elements!  Rather, when evaluating the product $\mathcal{M}F^*$ in the unitarity formula (\ref{cute_eigenvalue_equation}),
as shown in fig.~\ref{fig: two-loop cuts} one sees that $2{\to}3$ amplitudes and higher also contribute to the anomalous dimension of twist-two operators.
According to our main equation (\ref{cute_eigenvalue_equation}), anomalous dimensions are then
obtained by comparing this product $\mathcal{M}F^*$ with $-i(\e^{-i\pi D}-1)F^*$.
Using the dilatation operator $D$ given in eq.~(\ref{rg_equation}), one sees that at two-loops
this removes terms proportional to either the square of one-loop anomalous dimensions or to the one-loop $\beta$-function.

\def\middletension{0.8}

\subsection{General operators at one-loop level and the \texorpdfstring{$\mathcal{N}=4$}{N=4} spin chain}

We conclude this section by discussing general operators at one-loop level.
The main issue is the cancellation of infrared divergences for multiple external partons.
In principle, one could use again matrix elements of the stress tensor, but since they are coupling-constant suppressed this is not so convenient.
At one-loop, the tight structure of infrared divergences however makes this unnecessary.
The one-loop infrared anomalous dimension (defined by the renormalization group equation (\ref{eq: RG equation mu}) for the IR- and UV-renormalized form factor)
takes a very specific form in any gauge theory, see for example \cite{Sterman:2002qn,Becher:2009cu}:
\begin{equation}
\label{eq: general structure of one-loop IR}
 \gamma^{(1)}_\IR(\{p_i\};\mu)=\frac{g^2}{4\pi^2}\sum_{i<j} T_i^aT_j^a \log \frac{\mu^2}{-s_{ij}} + \sum_i \gcoll_i\,,
\end{equation}
where $T_i^a$ denotes the gauge-group generator acting on particle $i$. 
The fact that infrared divergences obey a renormalization group equation stems, of course,
from the general Wilsonian principle that disparate energy scales decouple;
we refer the reader to \cite{Feige:2014wja} for a recent explicit proof and further references.
Note that, in contrast to the ultraviolet case, infrared anomalous dimensions can depend explicitly on $\log\mu^2$ (at most linearly to any loop order),
reflecting double-logarithmic divergences from modes that are simultaneously soft and collinear.

The first term in \eqref{eq: general structure of one-loop IR}, coming from soft wide-angle radiation, can be identified with the integral over the
$1/(\sin^2\theta\cos^2\theta)$ term in eq.~\eqref{eq: Matrix element L},
which is the squared matrix element one would get from an integral over real radiation.
Therefore, the general one-loop dilatation operator (encoding all one-loop anomalous dimensions)
in an arbitrary gauge theory with matter contains a double-sum term, 
from the sum over unitarity cuts and soft contribution to the anomalous dimension, together with a single-sum term accounting for remaining hard-collinear divergences:
\begin{align}
 \label{eq: gamma general length}
  \gamma^{(1)}_\OO\l p_1,\ldots,p_n| \OO|0\r^{(0)} 
 =& -\frac{1}{\pi} \l p_1,\ldots,p_n| \sum_{i<j}\left(\mathcal{M}^{2{\leftarrow}2}_{ij}+\frac{2g^2T_i^a T_j^a}{\sin^2\theta\cos^2\theta}\right)\otimes \OO|0\r^{(0)}\\&{}{}+ \l p_1,\ldots,p_n| \OO|0\r^{(0)} 
\times\sum_{i=1}^n \gcoll_i\,.\nonumber
\end{align}
Here, $\MM_{ij}$ denotes the $2{\to}2$ amplitude acting on the final-state particles $i$ and $j$.
The two-body phase-space integral, which is represented by the convolution sign and
defined in eqs.~(\ref{eq: phase space})-(\ref{eq: one-loop measure}), is absolutely convergent for each term.

In QCD, matching and integrating the explicit expressions for the stress-tensor subtractions in eqs.~(\ref{eq: stress tensor QCD}) and (\ref{eq: fermion mass running}),
we get the one-loop collinear anomalous dimensions $\gcoll_g = -\frac{g^2b_0}{16\pi^2}$ and $\gcoll_\psi =-\frac{3g^2\CF}{16\pi^2}$,
again in agreement with standard results \cite{Becher:2009cu}.
The equality of the one-loop $\beta$-function and the collinear anomalous dimension (and the coefficient of $\delta(1{-}x)$ in eq.~(\ref{eq: DGLAP}))
can be attributed, in this framework,
to the simplicity of the Lagrangian form factor (\ref{eq: Matrix element L}), which exactly matches the soft (classical) infrared divergences.

An interesting special case of this formula is the planar limit of $\mathcal{N}=4$ super Yang-Mills.
In the planar limit, we consider single-trace operators, and $i$ and $j$ must be color-adjacent. Thus, we set
$j=i+1$ with $n+1\equiv 1$ identified following the cyclic invariance of the trace, and $T_i^aT_{i{+}1}^a\to -\tfrac{N_c}{2}$.
In this model, all states lie within one supermultiplet and are conveniently labelled
by polynomials in superspinors $\lambda_i,\tilde\lambda_i,\tildeeta_i$ as in eq.~(\ref{eq: Nair superamplitude}); summing over internal helicities, one finds that
the supermomentum-conserving $\delta$-function simply forces the $\tildeeta'$ to rotate like the $\tilde\lambda'$, so the right-hand side here will be evaluated with rotated superspinors
(\ref{rotation}).   Finally, the planar $2{\to}2$ amplitude is equal to the first term in eq.~(\ref{eq: Parke-Taylor formula}), which is larger than
the color-singlet amplitude (\ref{eq: Parke-Taylor formula_trace}) by a factor $\cos^2\theta$.
Substituting it into the above formula, we thus get\footnote{In the planar limit, it is conventional to not symmetrize in the two cut particles;
it can be verified that after symmetrization, using that $\cot\theta{+}\tan\theta{=}\frac{1}{\cos\theta\sin\theta}$,
the IR subtraction is exactly as in eq.~(\ref{eq: gamma general length}) with $\gcoll_{\mathcal{N}=4}=0$.}
\begin{align}
 \gamma^{(1)}_\OO \l 1, \ldots, n|\OO|0\r^{(0)} =
 \frac{4g^2N_c}{16\pi^2} \sum_{i=1}^n \int_0^{2\pi} \frac{\de\phi}{2\pi}\int_0^{\frac{\pi}{2}} \de\theta \cot\theta \left(
 \begin{array}{l}\phantom{+}
\l 1,\ldots, i,i{+}1,\ldots, n|\OO|0\r^{(0)} \\ -
\l 1,\ldots, i',(i{+}1)',\ldots, n|\OO|0\r^{(0)}\end{array}\right)\,.
\end{align}
This formula is precisely the one written down by Zwiebel for the one-loop dilatation operator in planar $\mathcal{N}=4$ SYM \cite{Zwiebel:2011bx},
 which in some way led to this work.

In various subsectors, the expression above reduces for example to the Hamiltonian of the integrable SU(2) or SL(2) Heisenberg spin chain,
revealing the integrability of the theory \cite{Minahan:2002ve,Beisert:2003yb}.
As far as we know, the original motivation of \cite{Zwiebel:2011bx} was based on symmetries: the one-loop dilatation operator and tree-level four-point $S$-matrix being both completely fixed by Yangian symmetry up to a multiplicative constant, they may be proportional to each other. 
This was then understood more directly from generalized unitarity \cite{Wilhelm:2014qua}.
In this paper, we have derived this formula using conventional unitarity and given a quantitative
extension to an arbitrary weakly coupled field theory, eq.~(\ref{eq: gamma general length}).

In large $N_c$-QCD, we thus expect that upon substituting the appropriate quark and gluon $2{\to}2$ tree amplitudes as in eq.~(\ref{mainresult}),
the formula will reproduce the one-loop dilatation operator from ref.~\cite{Beisert:2004fv}.
It would also be interesting to specialize the formula to the Standard Model and compare with the dimension-six anomalous dimensions, see for example
ref.~\cite{Alonso:2013hga};
certain qualitative features, such as zeros that are not obvious from Feynman diagrams, are nicely explained from unitarity and
on-shell tree-level helicity conservation rules \cite{Cheung:2015aba}.

\section{Length-changing effects and towards higher loops: Yukawa theory}
\label{sec: towards higher loops}

Let us now look at Yukawa theory, where we will encounter several new effects looking at operators of higher length and at higher loops.
These include mixing between operators of different lengths and the cancellation of logarithms between different cuts. 

For illustration, it will be sufficient to consider a theory with one real scalar and one Weyl fermion,
with interaction Lagrangian
\begin{equation}
 \label{eq: interaction Lagrangian}
 \mathcal{L}_{\rm int}=
 -\lambda\mathcal{O}_\lambda -y \mathcal{O}_y\quad\text{with}\quad\mathcal{O}_\lambda=\frac{1}{4!}\phi^4\quad\text{and}\quad\mathcal{O}_y=\frac{1}{2}(\psi\psi\phi+\text{h.c.})\,.
\end{equation}
The minimal form factors of the operators $\mathcal{O}_\lambda$ and $\mathcal{O}_y$ are 
\begin{equation} \label{eq: form factor}
 \l 1_\phi2_\phi3_\phi4_\phi| \OO_\lambda|0\r = 1\,,\quad
 \l 1_{\fermm}2_{\fermm}3_\phi| \OO_y|0\r = \l12\r\,,\quad
 \l 1_{\fermp}2_{\fermp}3_\phi| \OO_y|0\r = [12]\,.
\end{equation}
Correspondingly, the elemental scattering amplitudes are 
\begin{equation}
\label{eq: Yukawa theory elemental scattering amplitudes}
\begin{aligned}
  \mathcal{M}_{1_\phi2_\phi3_\phi4_\phi} &= -\lambda\,,\qquad
  \mathcal{M}_{1_{\fermm}2_{\fermm}3_\phi} &= -y\l12\r\,, \qquad
  \mathcal{M}_{1_{\fermp}2_{\fermp}3_\phi} &= -y[12]\,.
\end{aligned}
\end{equation}
One can check that the relative signs between the latter two amplitudes is consistent with unitarity,
so that $\l 1_\phi|2_\fermm3_\fermm\r\l2_\fermm3_\fermm|1_\phi\r\geq0$ as it should, using the crossing relation
(\ref{eq: crossing}).
Other amplitudes can be obtained using the factorization on poles:
\begin{equation}
\label{eq: Yukawa theory scattering amplitudes}
\begin{aligned}
  \mathcal{M}_{1_{\fermm}2_{\fermp}3_\phi4_\phi} &= y^2\left(\frac{\l13\r}{\l23\r}+\frac{\l14\r}{\l24\r}\right)\,, \\
  \mathcal{M}_{1_{\fermp}2_\fermp3_{\fermm}4_{\fermm}} &= y^2\frac{\l34\r}{\l12\r}\,, \\
  \mathcal{M}_{1_{\fermp}2_{\fermp}3_{\fermp}4_{\fermp}} &= 3y^2 \frac{[34]}{\l12\r}\,, \\
  \mathcal{M}_{1_{\fermp}2_{\fermp}3_\phi4_\phi5_\phi}&= y\lambda \frac{1}{\l12\r}-y^3\left( \frac{\l35\r}{\l13\r\l25\r} + \mbox{5 permutations of (345)}\right)\,.
\end{aligned}
\end{equation}
The signs of these amplitudes will be significant since in this approach they ultimately determine the sign of the anomalous dimension;
they are fixed for example by the factorization of trees $\l 12|\MM|34\r \to \l 12|\MM|i\r \frac{1}{\l12\r[12]} \l i|\MM|34\r$ in the limit where $p_i=(p_1+p_2)$ becomes null.

From the above scattering amplitudes and form factors, we will calculate the anomalous-dimension matrix
\be
 \left(\mu\frac{\partial}{\partial \mu} + \sum_{a=y,\lambda}\beta(a) \frac{\partial}{\partial a}\right)
  \left(\begin{array}{c} \OO_y\\ \OO_\lambda\end{array}\right)
=\left(\begin{array}{cc} \gamma_{yy}&\gamma_{y\lambda}\\ \gamma_{\lambda y}&\gamma_{\lambda\lambda}\end{array}\right)
\left(\begin{array}{c} \OO_y\\ \OO_\lambda\end{array}\right)\,.
\ee
From it one can then get $\beta$-functions,
using a generalization of the relation (\ref{eq:gamma vs beta function}) that we used in the Yang-Mills case.
We briefly recall its derivation \cite{KlubergStern:1974rs}.
First we note that we have normalized the operators so that their form factors (\ref{eq: form factor}) restricted to zero total momentum
are precisely the derivatives of the $S$-matrix with respect to the corresponding coupling:
\be
\mathcal{F}_a=-\frac{\partial}{\partial a} \mathcal{M} \,.
\ee
One now considers the RG equation for the UV (not IR)-renormalized amplitude and form factor (that is, contrary to what was done so far in this paper,
here we consider independent ultraviolet and infrared renormalization scales):
\be
  \left(\muUV\frac{\partial}{\partial \muUV} + \sum_{b=y,\lambda}\beta(b) \frac{\partial}{\partial b}\right)\mathcal{M}=0\,,
  \qquad
  \left(\muUV\frac{\partial}{\partial \muUV} + \sum_{b=y,\lambda}\beta(b) \frac{\partial}{\partial b}\right)\mathcal{F}_a=-\sum_{b=y,\lambda}\gamma_{ab}\mathcal{F}_b\,.
\ee
Deriving the first equation with respect to the coupling and comparing with the second equation gives the desired relation:
\be
 \frac{\partial}{\partial a}  \beta(b)=\gamma_{ab}\,,\qquad a,b= \lambda \, \mbox{ or }\, y\,. \label{eq: beta from gamma}
\ee

\subsection{IR structure and diagonal elements}

The diagonal (length-preserving) elements of the mixing matrix can be calculated straightforwardly using the by-now familiar procedure
of the preceding section: we act on form factors with the $2{\to}2$ tree amplitudes,
employing the stress tensor to remove the infrared (and collinear) contributions.  There are no new subtleties but if anything it is instructive to carry through this exercise.

Using the crossing relation (\ref{eq: crossing}) and the spinor products (\ref{eq: rotated spinor products}), the matrix elements
we will need are easily obtained from \eqref{eq: Yukawa theory scattering amplitudes}:
\be\begin{aligned}
 \l 1_\phi2_\phi| \mathcal{M}|1'_\fermm2'_\fermp\r&=-\l 1_\fermm2_\fermp|\mathcal{M}|1'_\phi2'_\phi\r^*= y^2\left(\frac{\cos\theta}{\sin\theta}-\frac{\sin\theta}{\cos\theta}\right)\e^{-i\phi}\,,
\\
  \l 1_\fermm2_\fermm |\mathcal{M}|1'_\fermm2'_\fermm\r&=\l 1_\fermm2_\fermp |\mathcal{M}|1'_\fermm2'_\fermp\r= -\l 1_\fermm2_\fermp |\mathcal{M}|1'_\fermp2'_\fermm\r=-y^2\,,
\\
 \l 1_\fermm2_\fermm |\mathcal{M}|1'_\fermp2'_\fermp\r &=-3y^2\,,
\qquad
 \l 1_\fermm2_\phi |\mathcal{M}|1'_\fermm2'_\phi\r= -y^2\frac{1+\cos^2\theta}{\cos\theta}\,. \label{eq: Yukawa rotated amplitudes}
\end{aligned}\ee
Multiplying the first by the tree form factor for the stress tensor to fermions, given in eq.~(\ref{eq: form factor of energy momentum YM with fermions}),
and performing the azimuthal integrals then gives
\begin{equation}\begin{aligned}
 2\gcoll_{\phi} &\equiv \frac{1}{\pi}
 \frac{\l 1_\phi2_\phi|\MM\otimes T^{\alpha\beta,\dot\alpha\dot\beta}|0\r^{(0)}}{\l 1_\phi2_\phi|T^{\alpha\beta,\dot\alpha\dot\beta}|0\r^{(0)}}
\\ &= \frac{1}{16\pi^2}\int_0^{\frac{\pi}{2}} 2\cos\theta\sin\theta\de\theta
\left(-\frac14\lambda \big(1+\cos(4\theta)\big)+ 6y^2 \cos^2(2\theta)\right)
= \frac{2y^2}{16\pi^2}
 \\
 \Rightarrow \gcoll_\phi&=\frac{y^2}{16\pi^2}\,,
\end{aligned}\label{eq: collinear phi}\end{equation}
where we have included also the scalar ($\lambda$ term) and antifermion (factor of 2) in the cut.
The $\lambda$ term integrates to zero: there are no one-loop IR divergences in pure $\phi^4$ theory, as expected.
Here we remark that, even though the form factors (\ref{eq: form factor of energy momentum YM with fermions}) contain many terms,
because they are fixed by symmetry the algebra is highly redundant and just the coefficient
of one term, for example $p_1^{\alpha\dalpha}p_1^{\beta\dbeta}$, is enough to determine the anomalous dimension.
Considering similarly the form factor for two fermions, we find
\begin{equation}\begin{aligned}
 2\gcoll_{\psi} &= \frac{1}{16\pi^2}\int_0^{\frac{\pi}{2}} 2\cos\theta\sin\theta\de\theta \left(
 2y^2\cos^2(2\theta)-y^2\cos(4\theta)
 \right)
= \frac{y^2}{16\pi^2}
\quad
 \Rightarrow \gcoll_\psi=\frac{\tfrac12y^2}{16\pi^2}\,,
\end{aligned}\end{equation}
where we have used the second, fourth and fifth of the amplitudes in eq.~(\ref{eq: Yukawa rotated amplitudes}).

With the infrared contributions under control, we can now calculate the diagonal matrix elements,
which is particularly trivial for the $\phi^4$ vertex correction since the four-scalar amplitude is just a constant so
each matrix element gives a factor $-\lambda/(16\pi^2)$:
\begin{align}
 \gamma^{(1)}_{\lambda\lambda} &= 4\gcoll_\phi
 -\frac{1}{\pi}
 \frac{\l 1_\phi2_\phi3_\phi4_\phi|\left(\MM_{12}+\MM_{13}+\MM_{14}+\MM_{23}+\MM_{24}+\MM_{34}\right) \otimes \mathcal{O}_\lambda|0\r^{(0)}}{\l 1_\phi2_\phi3_\phi4_\phi|\mathcal{O}_\lambda|0\r^{(0)}}
\nonumber \\
 &= \frac{4y^2}{16\pi^2} + \frac{6\lambda}{16\pi^2} \,.
 \label{one_loop_gamma_ll}
\end{align}
The $y^2$ term comes entirely from the collinear divergences, following eq.~(\ref{eq: gamma general length}).
In $\phi^4$ theory, it would be absent and, reassuringly, the relation (\ref{eq: beta from gamma}) would give
\be
 \beta(\lambda) = \frac{3\lambda^2}{16\pi^2} \qquad \mbox{($\phi^4$ theory)}\,,
\ee
which is of course the standard result.

For the Yukawa vertex renormalization, we have some angular integrals to do,
involving the third, sixth and two permutations of the seventh term in (\ref{eq: Yukawa rotated amplitudes}):
\begin{align}
 \label{one_loop_gamma_yy}
 \gamma^{(1)}_{yy} &=  2\gcoll_{\fermm}+\gcoll_\phi
 -\frac{1}{\pi}
 \frac{\l 1_{\fermm}2_{\fermm}3_\phi|\left(\MM_{12}+\MM_{13}+\MM_{23}\right) \otimes \mathcal{O}_y|0\r^{(0)}}{\l 1_{\fermm}2_{\fermm}3_\phi|\mathcal{O}_y|0\r^{(0)}}
 \\ &= \frac{2y^2}{16\pi^2}-\frac{1}{16\pi^2} \int_0^{\frac{\pi}{2}}2\cos\theta\sin\theta\de\theta\left(-4y^2-2y^2(1+\cos^2\theta)-2y^2(1+\sin^2\theta)\right)
 = \frac{12y^2}{16\pi^2} \,. \nonumber
\end{align}
Note that, even though the Yukawa interaction between identical fermions is often said to be attractive, the matrix elements here are mostly negative,
thus leading to a positive anomalous dimension and a positive contribution to the $\beta$-function.
The difference is because the conventional statement applies to non-relativistic massive fermions while we are looking here at 
the ultrarelativistic case where the amplitude involves a helicity flip and is quite different.

\subsection{Length-increasing effects: Yukawa coupling contributing to \texorpdfstring{$\phi^4$}{phi to the fourth}} \label{ssec: increasing}

We now turn to some novel effects not discussed earlier -- at one-loop we can also have length-increasing mixing,
for example between the operators $\mathcal{O}_\lambda$ and $\mathcal{O}_y$.
In terms of the unitarity method, this will involve the $2{\to}3$ amplitude acting on the minimal form factor,
as well as the $2{\to}2$ scattering acting on the \emph{non-minimal} form factor:
\begin{equation}
\label{eq: length increasing mixing}
\begin{aligned}
 \gamma^{(1)}_{y\lambda} &= -\frac{1}{\pi}
 \frac{\l 1_\phi2_\phi3_\phi4_\phi|\left(\MM^{2\leftarrow2}_{12}+\MM^{2\leftarrow2}_{13}+\MM^{2\leftarrow2}_{14}+\MM^{2\leftarrow2}_{23}+\MM^{2\leftarrow2}_{24}+\MM^{2\leftarrow2}_{34}\right) \otimes \mathcal{O}_y|0\r^{(0)}}{\l 1_\phi2_\phi3_\phi4_\phi|\mathcal{O}_\lambda|0\r^{(0)}}\\
 &\phantom{{}=} -\frac{1}{\pi}
 \frac{\l 1_\phi2_\phi3_\phi4_\phi|\left(\MM^{3\leftarrow2}_{123}+\MM^{3\leftarrow2}_{124}+\MM^{3\leftarrow2}_{134}+\MM^{3\leftarrow2}_{234}\right) \otimes \mathcal{O}_y|0\r^{(0)}}{\l 1_\phi2_\phi3_\phi4_\phi|\mathcal{O}_\lambda|0\r^{(0)}}\,.
\end{aligned} 
\end{equation}
The subscripts on the amplitude indicate the final state partons to which it is connected.
From the second line we get a rather simple $\lambda y$ term (see the last amplitude in eq.~(\ref{eq: Yukawa theory scattering amplitudes})),
but for $y^3$ contributions there will be a non-trivial interplay between the two lines.
However,  the sum of terms should give a polynomial since it is the form factor of a local operator.

\begin{figure}[htbp]
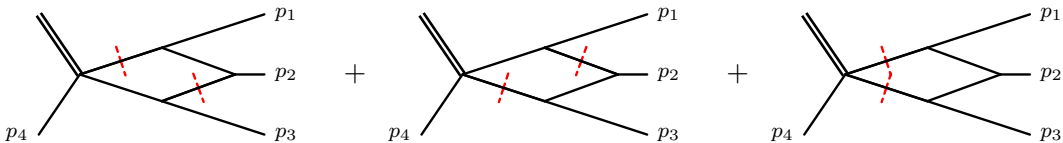

\centering
$
 \begin{aligned}
  \FDinline[boxplus,big,bslashdoublecut,fourlabels,labelone=p_1,labeltwo=p_2,labelthree=p_3,labelfour=p_4] 
 \end{aligned}
 \qquad
 +
 \quad
 \begin{aligned}
  \FDinline[boxplus,big,slashdoublecut,fourlabels,labelone=p_1,labeltwo=p_2,labelthree=p_3,labelfour=p_4] 
 \end{aligned}
 \qquad
 +
 \quad
 \begin{aligned}
  \FDinline[boxplus,big,doublecut,fourlabels,labelone=p_1,labeltwo=p_2,labelthree=p_3,labelfour=p_4] 
 \end{aligned}
$ 
\caption{Three cuts of the box integral among which logarithms cancel, see eq.~(\ref{eq: cancellation of logarithms}).}
\label{fig: cancellation in pictures}
\end{figure}

Anticipating that this will require cancellations, here we organize the terms into cuts of Feynman diagrams (since Feynman diagrams make locality manifest).
For example, consider the three cuts of the fermion box with on-shell scalars $p_1,p_2,p_3$ shown in fig.~\ref{fig: cancellation in pictures}.
The first cut is related to $\MM_{12}^{2\leftarrow 2}$ multiplied by the non-minimal form factor
\begin{equation}
\l 1_\fermp2_\fermm3_\phi4_\phi| \OO_y|0\r = -y\left( \frac{\l 13\r}{\l 23\r}+\frac{\l 14\r}{\l 24\r}-\frac{[23]}{[13]}-\frac{[24]}{[14]}\right).
\end{equation}
The sign of this expression can be verified by noting that for zero total momentum this is equal to minus
the $y$ derivative of the $2{\to}2$ $S$-matrix element given in the first line of eq.~(\ref{eq: Yukawa theory scattering amplitudes}).
The first cut comes from the first term in the amplitude (\ref{eq: Yukawa theory scattering amplitudes}) multiplied by the first term in the form factor:
\begin{equation}
 \l 1_\phi2_\phi| \mathcal{M}|1'_\fermp2'_\fermm\r_{\text{first term}}\l 1_\fermp'2_\fermm'3_\phi4_\phi| \OO_y|0\r_{\text{first term}}=y^3\frac{\l 2'1\r}{\l 1' 1\r}\frac{\l 1' 3\r}{\l 2' 3\r}\,.
\end{equation}
In the parametrization \eqref{rotation}, the phase-space integral reads
\begin{equation}
\begin{aligned}
 \int \frac{\de\Omega}{4\pi}\frac{\l 2'1\r}{\l 1' 1\r}\frac{\l 1' 3\r}{\l 2' 3\r}&=\int_0^{2\pi}\frac{\de\phi}{2\pi} \int_0^{\frac{\pi}{2}}\de\theta
\, 2\cos^2\theta \frac{\l13\r\cos\theta-\e^{i\phi}\l23\r\sin\theta }{\l13\r\sin\theta+\e^{i\phi} \l23\r\cos\theta}\,,
\end{aligned}
\end{equation}
where we have dropped the prefactor $-\frac{y^3}{16\pi^2}$ in the relation to the anomalous dimension.
A good way to perform the $\phi$ integral is as a contour integral over $z=\e^{i\phi}$ along the unit circle, which allows us to use Cauchy's residue theorem, obtaining
\begin{equation}
\label{eq: result double-cut two-particle channel}
\begin{aligned}
 - \int_0^{\frac{\pi}{2}}\de\theta \,2\cos^2\theta \left(\frac{\cos\theta }{\sin\theta }-\frac{1}{\cos\theta\sin\theta} \Theta\left(1-\left|\frac{\l 13\r\sin\theta}{\l 23\r\cos\theta}\right|\right) \right)=1+\log \frac{s_{23}}{s_{13}+s_{23}}\,.
\end{aligned}
\end{equation}
The step function $\Theta$ arises from whether the pole from the denominator is inside the unit circle.
The result from the double cut in the other two-particle channel can be obtained by replacing $1\leftrightarrow3$ in \eqref{eq: result double-cut two-particle channel}.
In this way, we have accounted for 2 out of the $2\times 6\times 2\times 4=96$ terms in the first line of eq.~(\ref{eq: length increasing mixing})
(one of the 2's is from exchanging $\fermm$ and $\fermp$).

We now consider the double cut in the three-particle channel.
We require both the three-point form factors in \eqref{eq: form factor} and the five-particle amplitude in the last line of \eqref{eq: Yukawa theory scattering amplitudes}.
As before, we will focus on one particular term corresponding to the third diagram of fig.~\ref{fig: cancellation in pictures}:
\begin{equation}
 \begin{aligned}
 \l 1_\phi2_\phi3_\phi| \mathcal{M}|1'_\fermm2'_\fermm\r_{\text{first $y^3$ term}}\, \l 1_\fermm'2_\fermm'4_\phi| \OO_y|0\r=
  \frac{\l 1 3\r}{\l 1' 1\r\l 2' 3\r}\l 1'2'\r \, .
 \end{aligned}
\end{equation}
We want to parametrize $p_1'$ and $p_2'$ as a rotation of suitable basis spinors.
In contrast to the cases above, we cannot simply take the base vectors to be external ones. Instead, we choose
\begin{equation}
 p_a=p_1 \frac{s_{123}}{s_{12}+s_{13}}\,, \qquad p_b=p_2+p_3-p_1 \frac{s_{23}}{s_{12}+s_{13}}\,,
\end{equation}
which are both on-shell and satisfy $p_a+p_b=p_1+p_2+p_3$. Corresponding spinors are
\begin{equation}
 \lambda_a=\lambda_1\sqrt{\frac{s_{123}}{s_{12}+s_{13}}}\, ,\qquad \lambda_b=([12]\lambda_2+[13]\lambda_3)\frac{1}{\sqrt{s_{12}+s_{13}}}\, .
\end{equation}
We then find
\begin{equation}
\begin{aligned}
\int \frac{\de\Omega}{4\pi}\frac{\l 1'2'\r\l 1 3\r}{\l 1' 1\r\l 2' 3\r}
  &=\int_0^{2\pi}\frac{\de \phi}{2\pi} \int_0^{\frac{\pi}{2}}\de\theta  \frac{2\cos\theta}{\sin\theta+\e^{i\phi}\cos\theta \frac{[12]\l23\r}{\l13\r\sqrt{s_{123}}}}\,,
\end{aligned}
\end{equation}
and, doing the $\phi$ integral again using Cauchy's theorem, we obtain
\begin{equation}
\label{eq: double-cut three-particle channel}
\begin{aligned}
 \int_0^{\frac{\pi}{2}}\de\theta  \,2\frac{\cos\theta}{\sin\theta}\Theta\left(1-\left|\frac{[12]\l 23\r\cos\theta}{\l 13\r\sqrt{s_{123}}\sin\theta}\right|\right) = \log \left(\frac{\left(s_{12}+s_{13}\right) \left(s_{13}+s_{23}\right)}{s_{12} s_{23}}\right)\, .
\end{aligned}
\end{equation}
Summing (\ref{eq: result double-cut two-particle channel}), its image under $1\leftrightarrow3$ and \eqref{eq: result double-cut two-particle channel} to get the three cuts in fig.~\ref{fig: cancellation in pictures} finally gives
\begin{equation}
\label{eq: cancellation of logarithms}
 \left(1+\log \frac{s_{23}}{s_{13}+s_{23}}\right)+ \left(1+\log \frac{s_{12}}{s_{12}+s_{13}}\right)+ \log \left(\frac{\left(s_{12}+s_{13}\right) \left(s_{13}+s_{23}\right)}{s_{12} s_{23}}\right)=2\,.
\end{equation}
As expected, the dependence on the kinematic variables has cancelled!
Restoring the factor $-y^3/(16\pi^2)$ and multiplying by 48 then gives the $y^3$ term in the anomalous dimension.
As already mentioned, there is also a simpler piece proportional to $y\lambda$, which comes only from the second line of eq.~(\ref{eq: length increasing mixing}) and involves the comparatively simpler
amplitude given in the last line of \eqref{eq: Yukawa theory scattering amplitudes}. In total, we thus get 
\begin{equation}
 \gamma_{y\lambda}=-\frac{96y^3}{16\pi^2}+\frac{8y\lambda}{16\pi^2}\,.
\end{equation}
Of course, the first term could have been obtained much more easily by extracting the ultraviolet-divergent part of the fermion box diagram. But this examples shows how,
through a non-trivial interplay between $S$-matrix elements and form factors responsible from the cancellation of logarithms (\ref{eq: cancellation of logarithms}),
the ultraviolet properties of the theories are also encoded in on-shell amplitudes with finite momentum.

Since all cuts ended up being computed by residues using Cauchy's formula, we can track the cancellations to the fact that the residues on triple cuts agree
regardless of the order in which the propagators are cut. 
Physically, this is a consequence of the factorization of amplitudes and form factors on their poles.
Understanding how to systematize such cancellations would be of great help for applications to the dilatation operator at lengths $\geq 3$ at higher-loops,
especially in gauge theories where the comparative simplicity of on-shell amplitudes adds a practical advantage to the method.

\subsection{Length-decreasing effects: a simple two-loop contribution}

Finally, we consider the length-decreasing mixing of $\mathcal{O}_\lambda$ into $\mathcal{O}_y$.
An important feature is that such mixing is not possible at one-loop: it would require a cut in a massless channel (with a $2{\to} 1$ amplitude on one side),
which of course is kinematically impossible.  
Therefore, the first length-decreasing effects occur at two-loops, through a $3{\to} 2$ amplitude integrated over a 3-particle cut.

There exist efficient modern techniques to deal with such two-loop cut integrals, notably by using integration-by-parts techniques and so-called reverse unitarity, see for example \cite{Anastasiou:2002yz}. 
Here, in line with previous examples, we adopt a low-tech approach and parametrize directly the angular integrals.
A price to pay is that we have to use different parametrizations for different terms in the amplitude.
We use eq.~(5.23) and preceding ones from \cite{Zwiebel:2011bx}:
\begin{equation}
\begin{aligned}
 {\lambda_1'}^\alpha&=\lambda_1^\alpha \cos\theta_2-\e^{i \phi } \lambda_2^\alpha \cos\theta_1 \sin\theta_2 \,,\\
 {\lambda_2'}^\alpha&=\lambda_1^\alpha \sin\theta_2 \cos\theta_3+\e^{i \phi } \lambda_2^\alpha \left(\cos\theta_1 \cos\theta_2 \cos\theta_3-\e^{i \rho } \sin\theta_1 \sin\theta_3\right)\,,\\
 {\lambda_3'}^\alpha&=\lambda_1^\alpha \sin\theta_2 \sin\theta_3+\e^{i \phi } \lambda_2^\alpha \left(\cos\theta_1 \cos\theta_2 \sin\theta_3+\e^{i \rho } \sin\theta_1 \cos\theta_3\right)\,.
\end{aligned}
\end{equation}
This has a simple physical interpretation in terms of collinearly splitting $p_2$ into two daughters with momentum fractions $\cos^2\theta_1$ and $\sin^2\theta_1$,
followed by applying the rotation (\ref{rotation}) on two different pairs.
A nice feature is that the propagators in the first $y^3$ term in eq.~(\ref{eq: Yukawa theory scattering amplitudes}) become elementary trigonometric functions:
\begin{equation}
\label{eq: amplitude 3 to 2}
\l 1_{\fermp}2_{\fermp}|\mathcal{M}^{2{\leftarrow}3}|1'_\phi2'_\phi3'_\phi\r\big|_{\text{first $y^3$ term}}
=-\frac{y^3}{\l12\r}(\e^{i \rho } \tan\theta_1 \cot\theta_2 \csc\theta_2 \cot\theta_3+\cot^2\theta_2+1)\,.
\end{equation}
In order to calculate the phase-space integral, we also require the measure factor, given as $(-1/\pi)$ times the phase-space volume.
It is given by 
\begin{equation}
\label{eq: measure 3 to 2}
 -\frac{s_{12}}{(4\pi)^4}\de \mu\quad \text{with} \quad\de\mu=2\sin\theta_1\cos\theta_1\de\theta_1\,4\sin^3\theta_2\cos\theta_2\de\theta_2\,2\sin\theta_3\cos\theta_3\de\theta_3 \frac{\de\rho}{2\pi}\frac{\de\phi}{2\pi}\,.
\end{equation}
In order to check the normalization of \eqref{eq: measure 3 to 2}, we compute
\begin{align}
  \int\de\mu&=\int_0^{\frac{\pi}{2}}2\sin\theta_1\cos\theta_1\de\theta_1\int_0^{\frac{\pi}{2}}4\sin^3\theta_2\cos\theta_2\de\theta_2\int_0^{\frac{\pi}{2}}2\sin\theta_3\cos\theta_3\de\theta_3 \int_0^{2\pi}\frac{\de\rho}{2\pi}\int_0^{2\pi}\frac{\de\phi}{2\pi}\nonumber\\
 &=1\,,
\label{eq: measure 3 to 2 integrated}
\end{align}
and we compare this to the discontinuity of the sunrise integral
\begin{equation}
\begin{aligned}
 -\frac{1}{\pi}\FDinline[sunrise,twolabels,labelone=p_1,labeltwo=p_2,triplecut] &= -\frac{2}{\pi}\Im\left[\frac{1}{(4\pi)^{4-2\peps}} \frac{\Gamma(1-\peps)^3\Gamma(1+2\peps)}{2\peps (1-2\peps)\Gamma(3-3\peps)}  (-s_{12}) \left(-\frac{s_{12}}{\mu^2} \right)^{-2\peps}\right]
 =-\frac{s_{12}}{(4\pi)^4} \,,
 \end{aligned}
\end{equation}
finding perfect agreement.
Integrating the amplitude \eqref{eq: amplitude 3 to 2} against the measure \eqref{eq: measure 3 to 2} with an additional symmetry factor of $\tfrac{1}{3!}$ for the three-particle phase space, we thus find 
\begin{equation}
 -\frac{s_{12}}{(4\pi)^4}\frac{1}{3!}\int\de \mu\l 1_{\fermp}2_{\fermp}|\mathcal{M}^{2{\leftarrow}3}|1'_\phi2'_\phi3'_\phi\r\big|_{\text{first $y^3$ term}}\l1_\phi'2_\phi'3_\phi'3_\phi|\OO_\lambda|0\r=-\frac{2y^3}{3!(4\pi)^4}\l 1_{\fermp}2_{\fermp}3_\phi| \OO_y|0\r\,.
\end{equation}
The contributions from the five permutations are identical. Finally, the contribution from the term in the amplitude proportional to $\lambda y$ can be integrated trivially, as it does not depend on the phase-space parameters.
Adding all seven terms, we find
\begin{equation}
 \gamma_{\lambda y}=-\frac{2y^3}{(4\pi)^4}+\frac{1}{6}\frac{y\lambda}{(4\pi)^4}\,.
\end{equation}

\subsection{Summary}

In total, we find
\begin{equation}
 \left(\begin{array}{cc} \gamma_{yy}&\gamma_{y\lambda}\\ \gamma_{\lambda y}&\gamma_{\lambda\lambda}\end{array}\right)
=
\frac{1}{16\pi^2} \left(\begin{array}{cc} 12y^2+O(y^4)& -96y^3+8y\lambda+O(y^5) \\ 
-\frac{2y^3}{16\pi^2}+\frac{1}{6}\frac{y\lambda}{16\pi^2}+O(y^5)& 6\lambda+4y^2+O(y^4)\end{array}\right)\,, \label{eq: Yukawa anomalous dimension result}
\end{equation}
where for simplicity we quote the errors in the technically natural power counting $\lambda\sim y^2$.
All of these entries are one-loop except for the lower-diagonal one, $\gamma_{\lambda y}$, for which we have included the two-loop contribution which is leading.

Integrating the relation (\ref{eq: beta from gamma}) between anomalous dimensions and $\beta$-function, $\gamma_{ab}=\partial_a \beta(b)$,
in particular yields the one-loop $\beta$-functions
\begin{equation}
 \beta^{(1)}(y) = \frac{1}{16\pi^2}\left( 4y^3\right),\qquad
 \beta^{(1)}(\lambda) = \frac{1}{16\pi^2}\left( -24y^4 +4y^2\lambda + 3\lambda^2\right)\,,
\end{equation}
which is the standard textbook result for the considered theory of one Weyl fermion and one real scalar; see for example \cite{Srednicki:2007qs}, up to minor modifications to reflect our matter content.
The computed two-loop entry also yields some simple two-loop contributions
\be
 \beta^{(2)}(y) = \frac{1}{(16\pi^2)^2}\left(-2y^3\lambda+ \frac{1}{12}y\lambda^2+ \mbox{undetermined terms proportional to $y^5$}\right)\,,
\ee
which can be compared for example with eq.\ (3.3) of \cite{Machacek:1983fi}, finding perfect agreement.
This demonstrates in a non-trivial way the correct handling of length-changing effects in the dilatation operator
by the proposed unitarity relation (\ref{cute_eigenvalue_equation}): $SF^*=\e^{-i\pi D}F^*$.

It is noteworthy that the relation between $\beta$-function and anomalous dimension is \emph{overconstrained}: the $4y^2\lambda$ term in $\beta(\lambda)$
is encoded in two different matrix elements of eq.~(\ref{eq: Yukawa anomalous dimension result}).
Since we obtained anomalous dimensions effectively as eigenvalues of the $S$-matrix, this must be viewed as a constraint satisfied by the $S$-matrix.
In fact, looking at the calculation, the $8y\lambda$ term in $\gamma_{y\lambda}$ is obtained from the $2{\to} 3$ amplitude in the last line of eq.~(\ref{eq: Yukawa theory scattering amplitudes}),
whereas the equivalent $4y^2$ term in $\gamma_{\lambda\lambda}$ comes from the collinear anomalous dimension (\ref{eq: collinear phi}), itself obtained from the $2{\to}2$ amplitudes acting on the stress tensor. It was not a-priori obvious why these $S$-matrix elements should be related, so it would be interesting to investigate such relations further.

It is interesting to see also that certain two-loop calculations in this section are actually simpler than one-loop calculations.
This is because, as presently formulated, calculating the anomalous dimension of high-length operators requires dealing with
a multi-scale problem (see eq.~(\ref{eq: cancellation of logarithms})) and so the number of legs has a strong impact on the complexity.

\section{Discussion and conclusion}
\label{sec: Summary}

In this paper, we have proposed a simple relation between the $S$-matrix of a theory at high energies and its dilatation operator:
\begin{equation}
\e^{-i\pi D} F^* = S F^*\,.  \label{eq: main equation conclusion}
\end{equation}
In essence, this states that the time evolution from asymptotic past to future, as encoded by the $S$-matrix, is equivalent to following a
half-circle generated by a complex scale transformation as shown in fig.~\ref{fig: half-circle}.
This means that {\it the dilatation operator is minus the phase of the $S$-matrix, divided by the circumference of the half-circle} ($\pi$).

At one-loop in Yang-Mills theory, this provides a surprisingly efficient way to calculate the $\beta$-function of the theory.
Starting with the famous Parke-Taylor tree-level amplitude for scattering four on-shell gluons  in eq.~(\ref{eq: Parke-Taylor formula}),
and performing elementary operations such as integrating over the two-body phase space, one reproduces the
famous result proportional to $-11\CA/3$ in section \ref{sec: YM eleven thirds}.  In particular, as usual with on-shell methods, only physical on-shell gluon states enter the calculation.
Furthermore, the sign is directly tied to the positive sign of the amplitude, itself stemming from the attractive force between opposite charges.
The extension to QCD including masses is discussed in subsection \ref{ssec: matter} and poses no significant problem.
We also found a correspondence between twist-two anomalous dimensions and the phase of $2{\to}2$ angular-momentum partial waves,
and obtained a novel formula, eq.~(\ref{eq: gamma general length}), for the one-loop dilatation operator in any gauge theory.
A pleasant feature is that the one-loop anomalous dimensions of all operators
are generated by the same building blocks, the $2{\to}2$ tree amplitudes of the theory.

Of course, in QCD the technology to calculate twist-two anomalous dimensions and $\beta$-functions is already very
well developed: for example, three-loop anomalous dimensions have been known for some time \cite{Vogt:2004mw,Moch:2004pa}
and the four- and even five-loop $\beta$-function are now known \cite{vanRitbergen:1997va,Czakon:2004bu,Baikov:2016tgj}.
On the other hand, for more general operators such as dimension-six operators in the Standard Model effective theory,
one-loop results have only been obtained recently \cite{Alonso:2013hga}.
An advantage of the present method is that it treats all operators of the theory on the same footing, which could help automation in this context.
In addition, certain qualitative features such as helicity selection rules are automatically manifest \cite{Cheung:2015aba}.
As mentioned in the main text, the present method could also be advantageous at higher loops in the context of theories with extended symmetries,
since the symmetries of the $S$-matrix are naturally maintained (including integrability in planar $\mathcal{N}=4$ SYM).

We have also investigated Yukawa theory at one-loop and beyond, confirming the general validity of the approach.
This is a phenomenologically important theory which
allows us to study effects which generically will be present at higher loops in any theory.
Of course, since on-shell amplitudes in this case are not simpler than the corresponding Feynman diagrams,
we did not expect a significant advantage to using this method.
The main new effect is that, while the $\beta$-function of Yang-Mills could be determined using an operator which decays
to two partons at tree level (the gluon density ${\rm Tr}\,[G^2]$), measuring the couplings in Yukawa theory requires more external legs interacting together, which makes the problem multi-scale and causes individual cuts to be more complicated.
The simplest case where this occurs is the length-increasing mixing at one-loop studied in section \ref{ssec: increasing}; here, logarithms cancel non-trivially between cuts (see eq.~\ref{eq: cancellation of logarithms}).
We expect such effects to be generic for higher-twist operators in any theory beyond one-loop,
and a formalism where such canceling transcendental functions could be discarded in individual cuts would greatly simplify calculations.
We note however that for twist-two operators and $\beta$-functions in QCD, the problem is always single-scale and such difficulties are absent.

In this work, we have taken the renormalization group equation as an input, but it is interesting to ask if it could be derived in an on-shell framework using physical
principles like unitarity of the $S$-matrix.  For example, there might be a recursive way to construct the scale dependence of the amplitudes and form factors on each side of a cut.
In general, the formalism exposes interesting relationships between form factors and the $S$-matrix, and it would be fascinating to study this interplay
in explicit examples at two loops and higher.

\section*{Acknowledgments}

We thank David McGady, Mike Trott and Florian Loebbert for discussion.
M.W.\ thanks Florian Loebbert, Christoph Sieg and Gang Yang for collaboration on a related project. 
M.W.\ was supported  in  part  by  DFF-FNU  through grant number DFF-4002-00037.
S.C.H.'s research was partly funded by the Danish National Research Foundation (DNRF91).
Both authors acknowledge the kind hospitality of NORDITA during the program ``Aspects of Amplitudes,'' where parts of this work were carried out.

\bibliographystyle{JHEP}
\bibliography{references}

\end{fmffile}
\end{document}